\documentclass[traditabstract]{aa}
\usepackage{epsfig}
\usepackage{hyperref}
\usepackage{amssymb,amsmath}
\usepackage{longtable}
\usepackage{natbib}
\usepackage{multirow}
\usepackage{tabularx}
\usepackage[farskip=0pt]{subfig}
\bibpunct{(}{)}{;}{a}{}{,} 

\newcolumntype{x}[1]{>{\centering\hspace{0pt}}p{#1}}
\newcommand{\planck}{\textit{Planck}}
\newcommand{\unit}{~\rm }

\begin{document}

\title{Effect of Fourier filters in removing periodic systematic effects from CMB data}
\titlerunning{Fourier filtering of periodic effects in CMB data}

   \author{
      F. de Gasperin\inst{1}\and
      A. Mennella\inst{2,3}\and
      D. Maino\inst{2}\and
      L. Terenzi\inst{4}\and
      S. Galeotta\inst{3}\and
      B. Cappellini\inst{2}\and
      G. Morgante\inst{4}\and
      M. Tomasi\inst{2}\and
      M. Bersanelli\inst{2}\and
      N. Mandolesi\inst{4}\and
      A. Zacchei\inst{3}
          }
   \authorrunning{F. de Gasperin et al.}
   
   \offprints{F. de Gasperin}

   \institute{
       Max Planck Institute for Astrophysics, Karl-Schwarzschild-Str. 1, 85741 Garching, Germany\and
       Universit\`a degli Studi di Milano, Dipartimento di Fisica, via Celoria 16, 20133 Milano, Italy \and
       INAF-OATs Trieste, Via Tiepolo 11, 34131 Trieste, Italy \and
       INAF-IASFBO, Via Gobetti 101, 40129 Bologna, Italy
   }

  \date{}

  \abstract{
    We consider the application of high-pass Fourier filters
    to remove periodic systematic
    fluctuations from full-sky survey CMB datasets.
    We compare the filter performance with destriping codes
    commonly used to remove the effect of residual $1/f$ noise from timelines.
    As a realistic working case, we use simulations of the typical \planck\ scanning strategy and \planck\
    Low Frequency Instrument noise performance, with spurious periodic fluctuations
    that mimic a typical thermal disturbance.
    We show that the application of Fourier high-pass filters in chunks always requires 
    subsequent normalisation
    of induced offsets by means of destriping. For a complex signal containing all the 
    astrophysical and instrumental components, the result obtained by applying filter and destriping in
    series is comparable to the result obtained by destriping only, which makes the usefulness of
    Fourier filters questionable for removing this kind of effects.
  }

  \keywords{Cosmology: cosmic background radiation - Cosmology: observation - Methods: data analysis}

\maketitle

%

\section{Introduction}
\label{sec:introduction}

  Controlling systematic errors is crucial to CMB anisotropy measurements, and will likely become one of the most important factors limiting the accuracy of future polarisation experiments performed with ultra-high sensitivity detector arrays \citep{2004astro.ph..2528M}. Second and third-generation CMB space missions (namely WMAP\footnote{http://map.gsfc.nasa.gov/} and \planck\footnote{http://planck.esa.int}) have been designed with tight systematic error control requirements calling for the development of spacecraft with stable thermal interfaces, optimized orbits, and scanning strategies and, in the case of \planck, advanced cryogenic systems for cooling high sensitivity differential receivers and bolometers with a high degree of stability. The sensitivity achievable by present and future experiments on CMB anisotropy and polarisation requires that residual systematics be controlled at the sub-$\mu$K level.

Cryo-coolers, in particular, can be the source of periodic systematic effects caused by fluctuations in the physical temperature of the receivers and the warm electronics. Even after optimal \textit{in-hardware} stabilization, a small level of residual fluctuations are generally present in the scientific data, with an impact that needs to be assessed to decide whether \textit{in-software} removal should be applied before science exploitation.

Because the thermal mass of both the satellite and the instrument damps high frequency temperature fluctuations, these effects are characterised by a frequency spectrum dominated by low frequencies, i.e. $\ll 1$~Hz, that propagate to the measured CMB maps leaving a signature after being reduced by redundant measurements performed in each pixel. From the point of view of scientific data quality, the amplitude of the angular power spectrum of this residual must be significantly smaller than the largest residual caused by $1/f$ noise fluctuations remaining after map-making. $1/f$ residual is present even when using optimal map-making approaches \citep[see, e.g.][]{ashdown2007} and appear as correlated structure on large scales below the level of white noise \citep{maino02a}.

Time-ordered data (TOD) from CMB experiments are generally processed before map-making to remove or reduce the contamination from spurious effects. Some codes, such as destriping \citep{kurki_suonio_2009, keihanen_2004, poutanen_2004} and Fourier filters \citep{tristram_2007, hivon2002} are able to achieve this  with no or minimal assumptions about the effect to be removed. They are robust, relatively easy to implement, and widely used as standard tools in CMB data analysis. Other codes make strong assumptions about the effect and use other data (e.g. from temperature and/or electrical sensors) to detect and remove the spurious signals. These can provide excellent results provided that accurate complementary information (e.g. from house-keeping telemetry such us the temperature recorded by a sensor on the instrument focal plane) is available. To remove thermal effects, for example, an effective use of temperature sensor data in non-blind codes calls for detailed knowledge about the amplitude and phase of thermal damping between the position of the temperature sensors and the detectors \citep[see, for example,][]{2009_LFI_cal_T3}.

In this paper, we consider the application of Fourier filters to CMB datasets, with particular reference to full-sky surveys performed from space. Fourier filters are well-suited to remove spurious effects with known spectral shape and have been applied to data acquired by sub-orbital missions. In the Boomerang experiment, for example, scan-synchronous effects were removed by filtering data with a high-pass filter with a cut frequency about seven times the scan frequency \citep{hivon2002, masi2006}.  The application of Fourier filters to large datasets such as those produced by space surveys, however, is not straightforward. Aggressive filtering would cut the sky signal on large angular scales that represent an essential part of full-sky CMB surveys. Furthermore, the large data stream size requires filters to be applied in chunks, introducing offsets that subsequently require normalisation.

The objective of this work is to analyse the application of high-pass filters to the removal of slow periodic effects from full-sky surveys datasets and compare the performance with destriping codes commonly used to remove the effect of residual 1/$f$ noise from timelines. Data used in this work consist of simulations representing the typical \planck\ scanning strategy and \planck-LFI\footnote{Low Frequency Instrument} noise performance \citep{2009_LFI_cal_M3,2009_LFI_cal_R2} with a periodic fluctuations representing a thermal perturbation in the 20 K stage. The spectrum of the fluctuations is based on laboratory tests on the 20 K cooler, while the amplitude is adjustable according to the need of the simulation. In any case, the fluctuations are not representative of the actual stability measured in-flight. We stress therefore that our analysis does not provide an assessment of the thermal systematic errors in \planck, but a tool to evaluate the effectiveness of Fourier filters in removing periodic spurious fluctuations in CMB data.
  

\section{Residual of periodic systematic effects in CMB maps}
\label{sec:periodic_systematic_effects_overview}
  We introduce now some general considerations about the propagation
of periodic fluctuations from measured time streams to CMB maps. Further details about this analytical description and a comparison with simulations performed in the context of \planck\ can be found in \citet{mennella02}.

In what follows, we assume that each pixel of a sky ring is measured $N$ times before the optical
axis is repointed by an angle $\theta_{\rm rep}$. If we consider a periodic fluctuation of general shape in the detected signal ($\delta T$), we can expand it as the Fourier series $\delta T = \sum_{j=-\infty}^{+\infty}A_j \exp(i 2\pi f_j
t)$, where $f_j$ represents the various fluctuation frequency components.

To estimate how the amplitude of each component is reduced by the measurement redundancy provided by the scanning strategy, we divide the frequency spectrum of the systematic effect into two regions with respect to the scanning frequency, $f_{\rm scan}$: (i) the \textit{low} frequency region, with $f_j < f_{\rm scan}$, and (ii) the \textit{high} frequency region, with $f_j \ge f_{\rm scan}$.

In the low frequency region, each harmonic of amplitude $A_j$ will be damped by the measurement redundancy by a factor proportional to $\sin(\pi f_j / f_{\rm scan})$. In the high frequency region, instead, we differentiate between scan-synchronous (i.e. $f_j = k\, f_{\rm scan}$) and scan-asynchronous (i.e. $f_j \ne k\, f_{\rm scan}$) fluctuations. The first type will not be damped, as they are practically indistinguishable from the sky measurement; the second ones, instead, will be damped by the number of measurements for that pixel, that is a factor of the order of $N\times \theta_{\rm pix} / \theta_{\rm rep}$ where $\theta_{\rm pix}$ is the pixel angular dimension.

If we also consider the additional reduction provided by the application of removal algorithms to the TOD, and denote with $F_j$ the additional damping of each frequency component, then the final peak-to-peak effect of a generic signal fluctuation $\delta T$ on the map can be estimated as

\begin{eqnarray}
    &&\langle\delta T^{\rm p-p}\rangle_{\rm map} \sim\nonumber \\
    &&2\left[
    \frac{1}{N\times \theta_{\rm pix}/\theta_{\rm rep}}\left(
    \sum_{f_j < f_{\rm scan}}
    \left|\frac{A_j/F_j}{\sin(\pi f_j / f_{\rm scan} )}\right| + \right. \right.\\
    &&+\left.\left.\sum_{f_j > f_{\rm scan},f_j \neq f_{\rm scan}}
    \left|A_j/F_j\right| 
    \right)+
    \sum_{f_j = k\, f_{\rm scan}} A_j\right]. \nonumber
    \label{eq:p2p_map_with_destriping}
\end{eqnarray}

%
%
%
\section{Map-making and removal approaches}
\label{sec:removal_approaches}

  Map-making is the process that combines the satellite pointing information and the instrument TODs into a map \citep{wright1996a,tegmark1997b,stompor2002}. Simple maps can be obtained by simply phase-binning data belonging to the same sky ring and then by averaging binned data that are observed in the same sky pixel. This method produces a raw map that is usually affected by spurious signatures caused by 1/$f$ noise, long-term drifts and slow periodic effects. More sophisticated methods can be divided into \textit{optimal}, least squares map-making \citep{wright1996b, tegmark1997b, tegmark1997a, borrill2001, natoli2001, dore2001}, and approximate methods, such as \textit{destriping} \citep[e.g.][]{delabrouille1998, burigana1999, maino1999, revenu2000, keihanen_2004, kurki_suonio_2009}.

Optimal methods require the inversion of large matrices or the use of iterative algorithms to avoid direct matrix inversions.  Destriping algorithms, however, do not yield optimal maps but are much less demanding on both memory and CPU \citep{poutanen_2004, stompor2004,efstathiou2005,efstathiou2007} and have been successfully applied to CMB data from the \planck\ space mission \citep{zacchei2011}. In this work, we produced maps using the \planck-LFI destriping code, which is briefly described in the next section.

\subsection{Destriping}
\label{sec:destriping}

The \textit{destriping} technique has been developed to reduce the effect of $1/f$ noise fluctuations. The main requirement is that the knee frequency, $f_{\rm knee}$, must not be much higher than the scan frequency, $f_{\rm scan}$; previous simulation works performed in the context of \planck\ \citep{kurki_suonio_2009} have shown that destriping is effective for knee frequencies up to 0.1~Hz, i.e. up to about six times the spin frequency.

According to this approach, the contribution from $1/f$ noise or other slow fluctuations is approximated as a constant over a certain length of time. The code then estimates these constant baselines and removes them from the TOD. The baseline length (that was fixed to 60 seconds in our simulations) can be chosen
to optimise the cleaning, according to the time-scale of the systematic effect.

To identify the optimal baselines, ring crossings (i.e. common pixels observed from different scan circles) are searched into every baseline-long set of data. We indicate with $T_{i,l}$ and $n_{i,l}$ the observed signal and the white noise level for the pixel in the $i$-th row (circle) and $j$-th column (sampling along the circle) in matrices $T$ and $n$, respectively. Baselines $A_i$ are then recovered by minimising the quantity
\begin{equation}
   S = \sum_{\rm crossings} \left[ \frac{ \left[ \left( A_i-A_j \right) - 
   \left( T_{i,l} - T_{j,m} \right) \right] ^2 }{ n^2_{i,l} - n^2_{j,m} } \right],
   \label{eq_destriping}
\end{equation}
where the sum is over all the pairs of pixels present in the two different baseline-long set of data. Baselines are recovered by solving a set of linear equations and then can be easily subtracted from the data.

Because destriping has also proven effective in removing slow periodic systematic effects \citep{zacchei2011}, we also compare its performance with that of the high-pass filter.

\subsection{TOD processing before map-making}
\label{sec:tod_processing}
    Although destriping can reduce the impact of low frequency modes, greater suppression may be achieved using additional algorithms that process TODs prior to map-making and can operate in ``blind'' or ``non-blind'' modes.  
    Blind codes make no or little assumptions about the physical and/or statistical properties of the signals to be removed and use temporal information in the data. When combined with map-making, the spatial information (i.e. the corresponding pixel position for an observed data-point on the final map) can also contribute to the final suppression level of the systematic effect.
    
    Non-blind codes use the available house-keeping (H/K) data, providing auxiliary information such as currents and voltages to the various electronic devices, temperature sensors data etc. 
    A simple example of non-blind analysis is the correlation between instrument output and H/K data. These methods are implemented by correlating the relevant H/K data with the radiometric output also using, where available, transfer functions obtained from dedicated tests.
    Another non-blind approach that has been studied within the \planck\ collaboration is one based on neural network algorithms \citep{maris2004_removal_thermal_effects}.
    In this case, the neural network is ``trained'' using simulated sensor data together with radiometer output obtained via analytical transfer functions. As usual, after a ``training'' period, the network is fed with real instrument output and H/K data.
    Both methods clearly require H/K data from many sensors placed very close to the detectors on the focal plane. However, this is not always possible because of constraints on the number and position of the sensors in the instrument focal plane, which often limit the effectiveness and reliability of non-blind approaches.
%
%
%
\section{Data and procedures}
\label{sec:data_procedures}

  We describe now the procedure we used to
simulate the time-streams and temperature anisotropy maps
containing astrophysical and instrumental signatures.
We also describe the high-pass filter used to remove the low frequency
components of the instrumental periodic fluctuations.

The simulations were executed considering the \planck\ scanning strategy and, in particular, instrumental
parameters typical of the \planck-LFI 30 GHz receivers. A template for the periodic spurious fluctuation
was provided by the expected temperature fluctuations in the 20~K LFI focal plane unit.

\subsection{Scanning strategy and sampling}
\label{sec:scanning_strategy_and_sampling}

    All the simulations were performed using the so-called ``cycloidal'' scanning strategy.
    According to this scheme, the satellite orbits around the L2 Lagrangian point of the Sun-Earth
    system and spins around its axis (at 1 rpm in our simulations), 
    which is pointed towards the Sun with the solar panels
    keeping the payload in shade. 
    The telescope points at an angle of $85^\circ$ with respect to the spin axis and sweeps the sky in near-great circles.
    The spin axis of the satellite also performs a 
    slow circular path around the anti-solar axis, which is equivalent to a cycloidal path as a function of ecliptic longitude.

    With a repointing of $2'$ every 48 minutes, the entire sky is observed by all detectors in the focal plane
    in about 7 months. A small hole\footnote{226 pixels in a map with $N_{\rm pixel}$ = 786432, that is ~0.03\%.}
    near the south pole was filled during the data-reduction procedure with values $\Delta T=0\unit K$.

    All data-streams were produced with a sampling frequency of 32.5 Hz, which corresponds to that of the 
    $30\unit GHz$ \planck-LFI radiometers and assuming that no gaps are present in the data streams.

\subsection{Time streams}
\label{sec:time_streams}

    \subsubsection{Astrophysical signal}
    \label{sec:sky_timestream}
        The sky emission template was provided by the \textit{Planck Sky Model}~(PSM), 
        which consists of a set of tools that simulate the whole sky emission in the frequency range $30-1000\unit GHz$. 
        In our study, we used a sky signal composed of the CMB and the diffuse foreground emission.

	The CMB was generated starting from the temperature power spectrum and fitting five years WMAP temperature data \citep{hinshaw2009},
        while for the foreground emissions we assumed three galactic components: 
        thermal dust, synchrotron, and free-free. 
        The SZ effect and strong point sources were not included because their 
        effect on the long-period variations are negligible. 

        A main beam with $\theta_{\rm FWHM} = 32.4$~arcmin
        was convolved with the sky signal to obtain a realistic data-stream. The main beam was simulated
        with the software GRASP using measured feed-horn beam patterns and a detailed model
        of the \planck\ telescope, without considering the beam far side lobes. 
        For more details about the LFI feed-horn patterns and estimated
        beams in the sky, the reader can refer to \citet{2009_LFI_cal_O1} and \citet{2009_LFI_cal_M5}.

    \subsubsection{Instrumental noise}
    \label{sec:instrumental_noise}
    
        The instrumental noise was simulated as a combination of white noise plus a 
        $1/f^\alpha$ component with a power spectrum of the form
        \begin{equation}
            P(f) = \sigma^2 \left[ 1+\left( \frac{f_{\rm knee}}{f} \right)^\alpha \right],
            \label{eq_one_over_f_power_spectrum}
        \end{equation}         
        where $\sigma^2$ is the white noise variance, $f_{\rm knee}$ is the frequency
        at which white noise and 1/$f^\alpha$ contribute equally in power and $\alpha$ is the exponent 
        in the power law that is usually in the range $0.5\lesssim\alpha\lesssim 2$.
        
        The presence of $1/f^\alpha$ noise in CMB data acquired by coherent
        receivers such as those used in \planck-LFI \citep{2009_LFI_cal_M2} and WMAP \citep{bennett03a}
        is caused by gain and noise temperature fluctuations
        in the high electron mobility transistor (HEMT) amplifiers. These fluctuations can be highly suppressed
        in hardware
        by adopting efficient differential receiver schemes \citep[see, e.g.:][]{seiffert02, mennella03}.

        Noise time-streams were simulated using the inverse Fourier transforming equation~(\ref{eq_one_over_f_power_spectrum})
        with the noise parameters reported in Table~\ref{tab:noise_parameters}\footnote{These parameters are representative
        of \planck-LFI 30 GHz receivers noise performances as measured during the instrument ground test
        campaign \citep{2009_LFI_cal_M3, 2009_LFI_cal_R2}}. 
        In particular, the noise spectral density $\delta T_{\rm 1sec-ANT}$ is related to the white noise
        variance $\sigma^2$ by the equation $\sigma = \delta T_{\rm 1sec-ANT} / \sqrt{\tau_{\rm int}}$, 
        where $\tau_{\rm int} = 1/f_{\rm samp}$ is the sample integration time.
        
        \begin{table}[!htb]
            \centering\caption{Typical noise parameters for \planck-LFI $30\unit GHz$ receivers}
            \label{tab:noise_parameters}
            \vspace{0.2cm}
            \footnotesize
            \def\arraystretch{1.5}
            \begin{tabular}{cccc}
                \hline\hline
                $f_{\rm knee}$ & $\alpha$ & $\delta T_{\rm 1sec-ANT}$ & $f_{\rm samp}$\tabularnewline
                \hline
                0.008 Hz & 0.69 & $2.629 \cdot 10^{-4}\unit K\, s^{1/2}$ & 32.5\,Hz\tabularnewline
                \hline
            \end{tabular}
        \end{table}

        \subsubsection{Periodic systematic effect}
        \label{sec:periodic_systematic_effect}

        Periodic, ``slow'' spurious fluctuations in cryogenic microwave receivers are often
        caused by temperature instabilities in the cryogenic system that propagate through the mechanical
        structure and couple with the measured signal. In coherent receivers, for example, this coupling
        is provided by the amplifier gain and noise temperature that oscillate synchronously with the 
        environment physical temperature \citep{2009_LFI_cal_R6}.

        In our work, we simulated a realistic periodic effect starting from the expected stability behaviour of the 
        \planck\ sorption cooler \citep{Bhandari04:planck_sorption_cooler,2009_LFI_cal_T0}, 
        which is the main cryogenic system on board the \planck\ 
        satellite providing a 20~K stage for the LFI receivers and a 18~K precooling stage
        for the HFI 4~K cooler. The main frequencies
        of the temperature fluctuations depend on the (programmable) absorption-desorption period of both each single
        compressor and the whole 6-compressor assembly. We chose a
        typical value of $940\unit s$ for the single compressor period which leads to $5640\unit s$ 
        for the main period of the whole assembly. The design temperature stability at the cooler cold
        end is $\sim 0.1\unit K$ peak-to-peak.
        
        Details of the thermal damping provided by the instrument structure and of the radiometric coupling
        functions are beyond the scope of this paper and not discussed here. Interested readers can find
        details in \citet{2009_LFI_cal_R6} and \citet{2009_LFI_cal_T3}. 
        
\begin{figure*}
            \includegraphics[width=9cm]{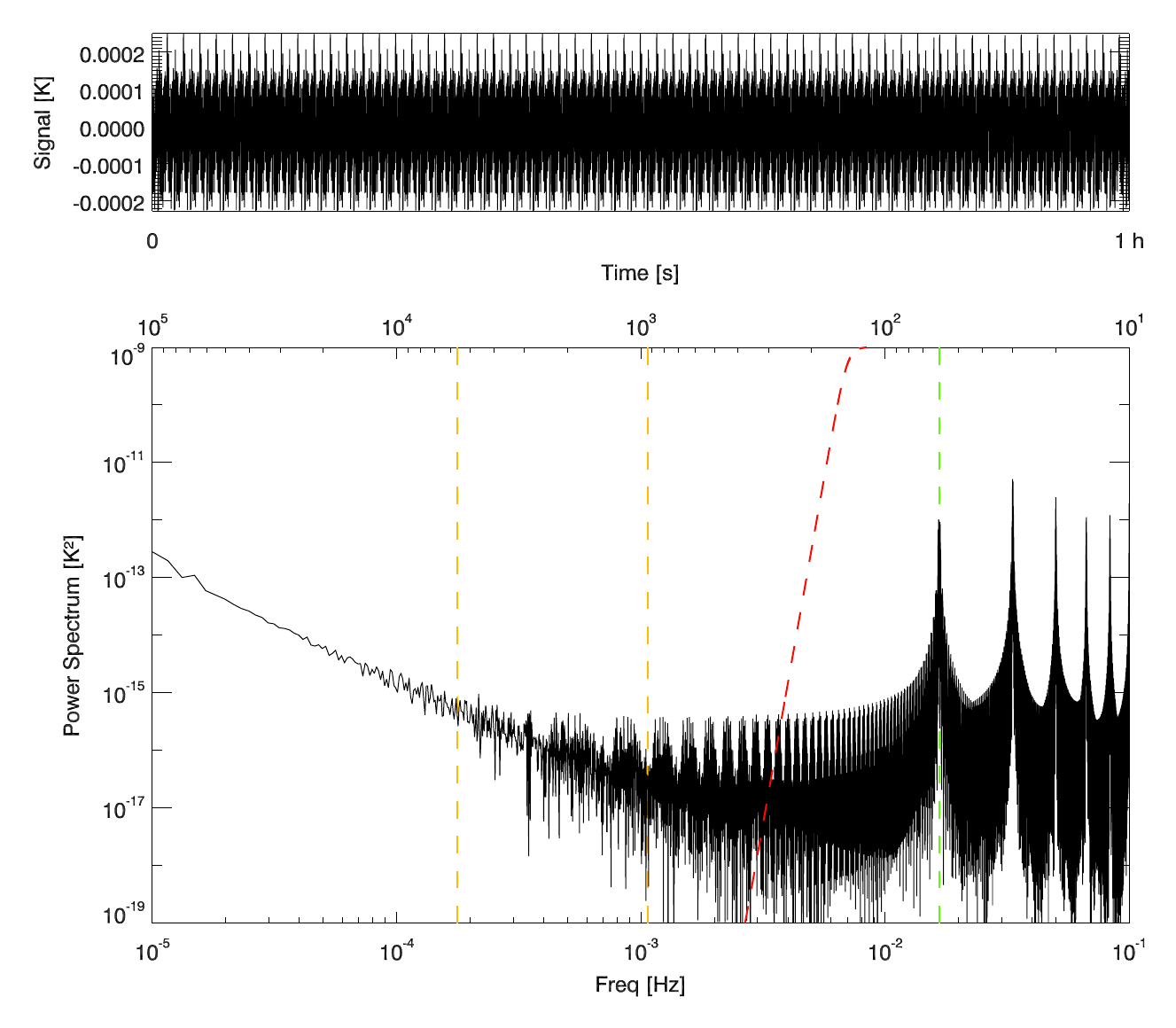}
            \includegraphics[width=9cm]{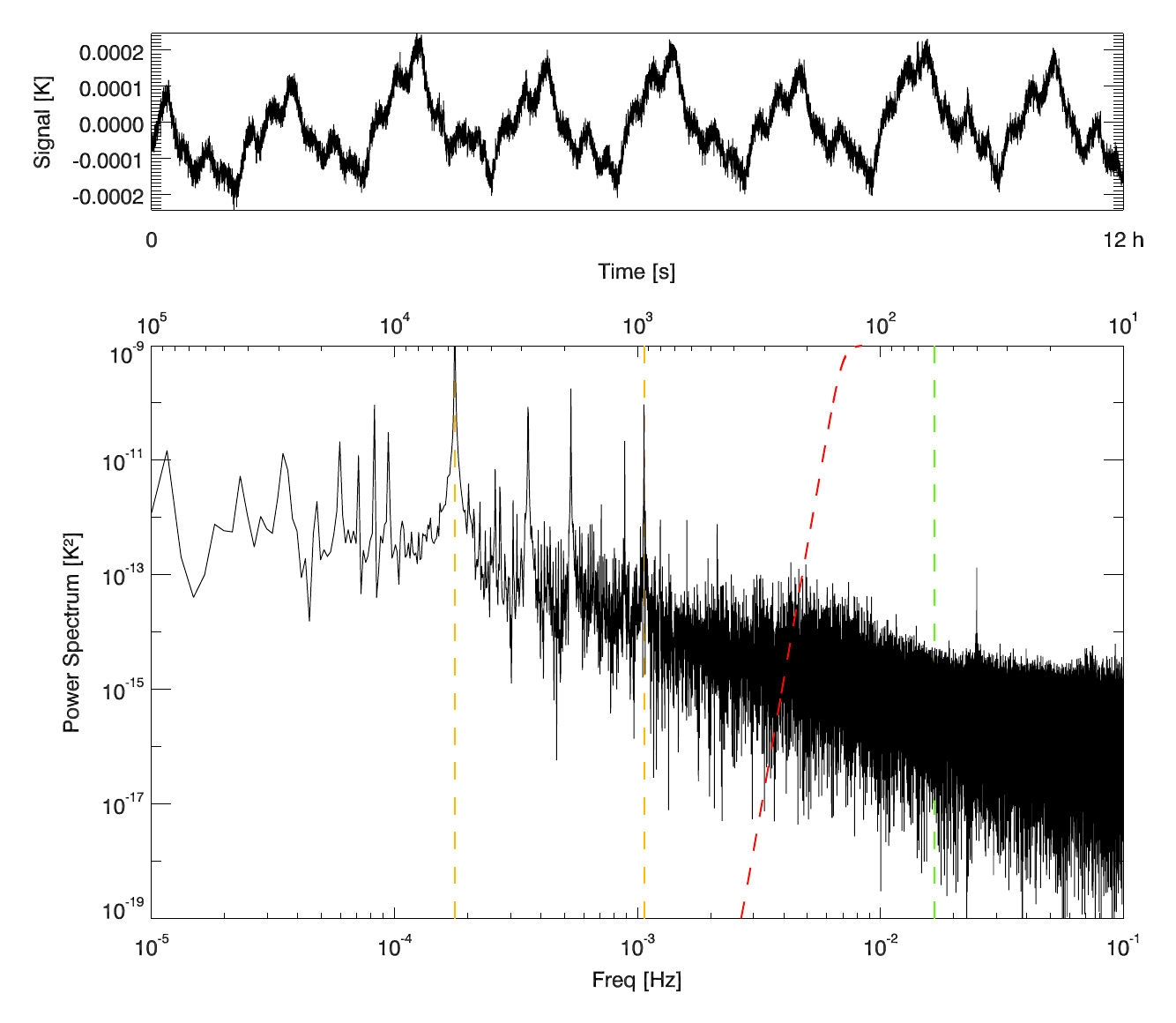}
            \caption{The astrophysical signal (left) and the periodic systematic effect (right) in the simulated data-streams. The top panels of the two
            figures present the signals in the time domain, while the bottom panels show their power spectra. 
            The green line is located at the frequency of $1/60\unit Hz$, 
            while the two yellow lines 
            indicate the two main frequencies of the periodic effect. The red dashed line shows the shape of the 
            high-pass filter; the filter curve is not to scale, its true value on the y-axis ranges from 0 to 1.}
            \label{fig:tods}
        \end{figure*}

\begin{figure*}[t]
            \begin{center}
                \includegraphics[width=14cm, angle=180]{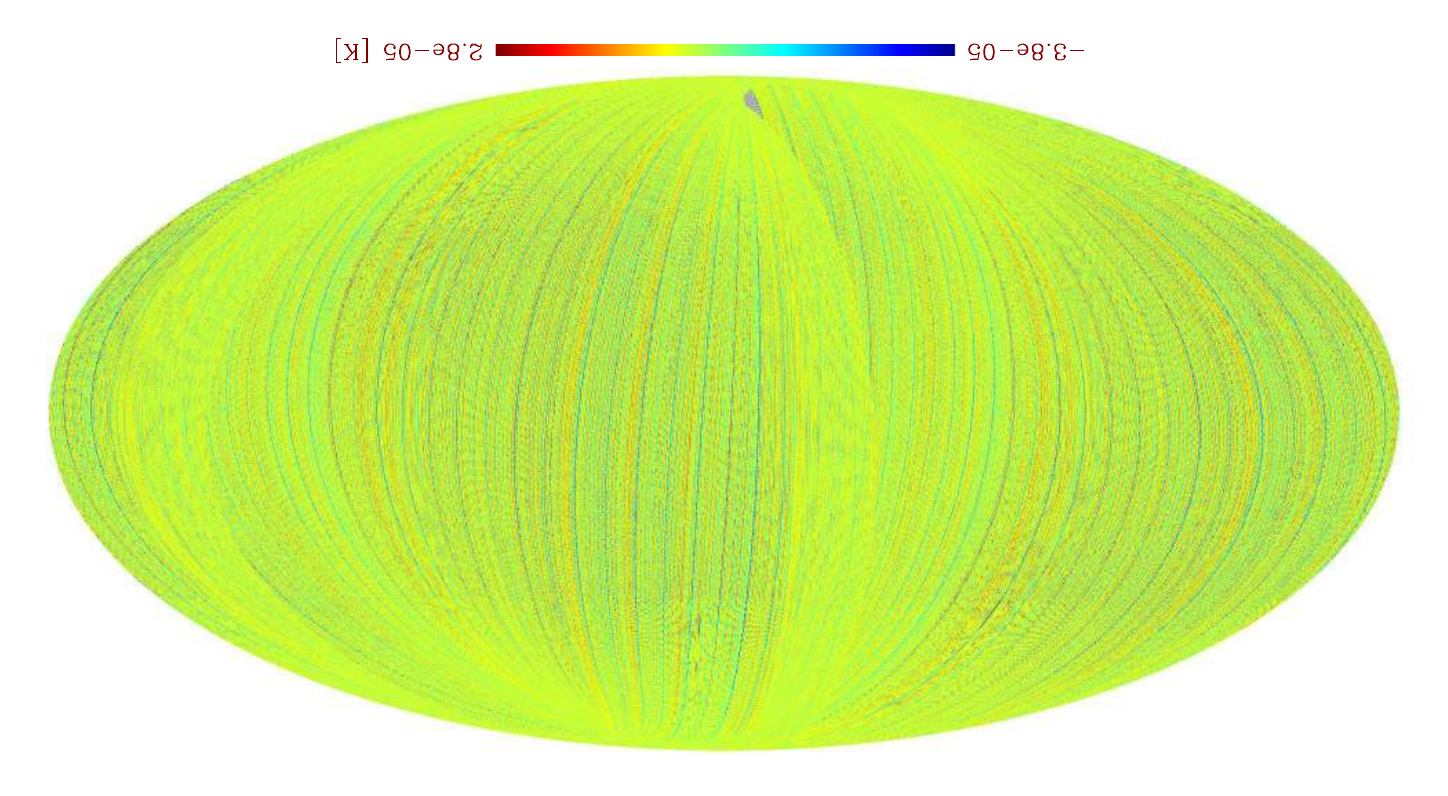}
                \caption{Map of the periodic systematic effect for a $30\unit GHz$ radiometer after a 
                survey of 7 months showing structures caused by the slow
                signal periodic fluctuations. 
                }
                \label{fig:map256-staticR_cooler_LFI-27a_1survey}
            \end{center}
        \end{figure*}

        In the right panel of Fig.~\ref{fig:tods}, we show an example of the periodic effect
        in the receiver output in the
        time and frequency domains. The peak-to-peak effect in antenna temperature is $\sim 4$~mK and 
        the two yellow lines in the spectrum highlight the main frequency peaks of
        the systematic effect.
        In Fig.~\ref{fig:map256-staticR_cooler_LFI-27a_1survey}, the same effect is shown when projected onto a map.
        The visible stripes are the signature of the systematic effect that has been damped 
        by the redundancy in the scanning strategy (see Sect.~\ref{sec:periodic_systematic_effects_overview})
        to a peak-to-peak of $\lesssim 70\unit \mu K$.

\subsection{High-pass filter}
\label{sec:filter}

    The power spectra of both the astrophysical signal and the periodic effect shown in Fig.~\ref{fig:tods}
    highlight their different distribution in the frequency domain. 
    Because of the scanning strategy, the astrophysical signal lies mainly at the frequency of $1/60\unit Hz$ and overtones;
    the periodic effect, instead, is characterised by fluctuations at lower frequencies. 
    A high-pass filter appears a natural solution to remove the systematic effect without affecting
    the astrophysical component.

    The filter action $H(f)$ in the frequency domain, can be described by the equation
    \begin{equation}
        [\hat{S}(f)]_{\rm HPF} = H(f) \cdot \left( [\hat{S}(f)]_{\rm sky} + 
        [\hat{S}(f)]_{\rm P} + [\hat{S}(f)]_{1/f} \right),
        \label{eq_filter_action}
    \end{equation}
    where $[\hat{S}(f)]_{\rm HPF} $ is the spectrum of the filtered signal,
    $[\hat{S}(f)]_{sky}$, $[\hat{S}(f)]_{P}$, and $[\hat{S}(f)]_{1/f}$ are the Fourier transforms
    of the astrophysical component, the periodic systematic effect, and the instrument noise. 
    
    The high-pass filter used in our work 
    is a simple real Butterworth filter \citep{butterworth} of the form
    \begin{equation}
        H(f) = \frac{\left(f/f_{\rm cut}\right)^n}{1+\left(f/f_{\rm cut}\right)^n},
        \label{eq_butterworth}
    \end{equation}
    where $f_{\rm cut}$ represents the filter cut-off frequency and $n$ the slope.
    This filter removes almost all the signal at frequencies $f < f_{\rm cut}$, leaving the complementary range of frequencies nearly untouched. Because the Fourier transform is linear, the filter can be applied to 
    the different signals separately to remove the systematic effect and 
    leave the astrophysical signal unchanged.

    A C++ code was developed and run on a 32-node cluster of 
    Intel Xeon $3.00\unit GHz$ processors with $2.00\unit GB$ SDRAM each; on this system, the code can
    process a seven-month dataset for a single detector in few minutes.
    
    Our first study has been a sensitivity analysis to determine the optimal cut frequency, $f_{\rm cut}$, 
    and the \textit{filter step}, $N$, i.e. the length of data that the filter will 
    process at each step. This last parameter, in particular, is required to avoid
    cutting the sky signal on large angular scales. Moreover, several months of data cannot be filtered
    in a single step because of memory limitations. The filter slope, $n$, was fixed in all our
    simulations at $n=24$, a value that makes the filter very steep yet avoids ringing side lobes that would be caused
    by a step function.

    In the sensitivity analysis, we optimised each parameter independently, while keeping the other two fixed at a reference value. We first generated a data-stream with all the components described in Sect.~\ref{sec:time_streams}
    and with a periodic effect amplitude 100 times larger than the one shown in Fig.~\ref{fig:tods}.
    We then calculated residual maps by subtracting the astrophysical signal map from the cleaned all-components map, i.e.
    \begin{equation}\label{eq:residual}
        M_{\rm(residual)} = \left[ M_{\rm(sky+systematics)}\right]_{\rm HPF} - M_{\rm(sky)},
    \end{equation}
    where $M_{\rm(sky)}$ is the map containing only the sky signal,
    while $\left[ M_{\rm(sky+systematics)}\right]_{\rm HPF}$ is the map of all the components
    (i.e. astrophysical, instrument noise and the periodic effect) after high-pass filtering.
    We finally compared the angular power spectra, $C_\ell^{\rm res}$,
    calculated\footnote{The angular power spectra were computed
    using the HEALPix (\url{http://healpix.jpl.nasa.gov)} utility \texttt{anafast}.}
    from residual maps obtained with different values of the considered filter parameter.

\begin{figure}[t]
        \begin{center}
            \centering
            \resizebox{\columnwidth}{!}{\includegraphics[width=9cm]{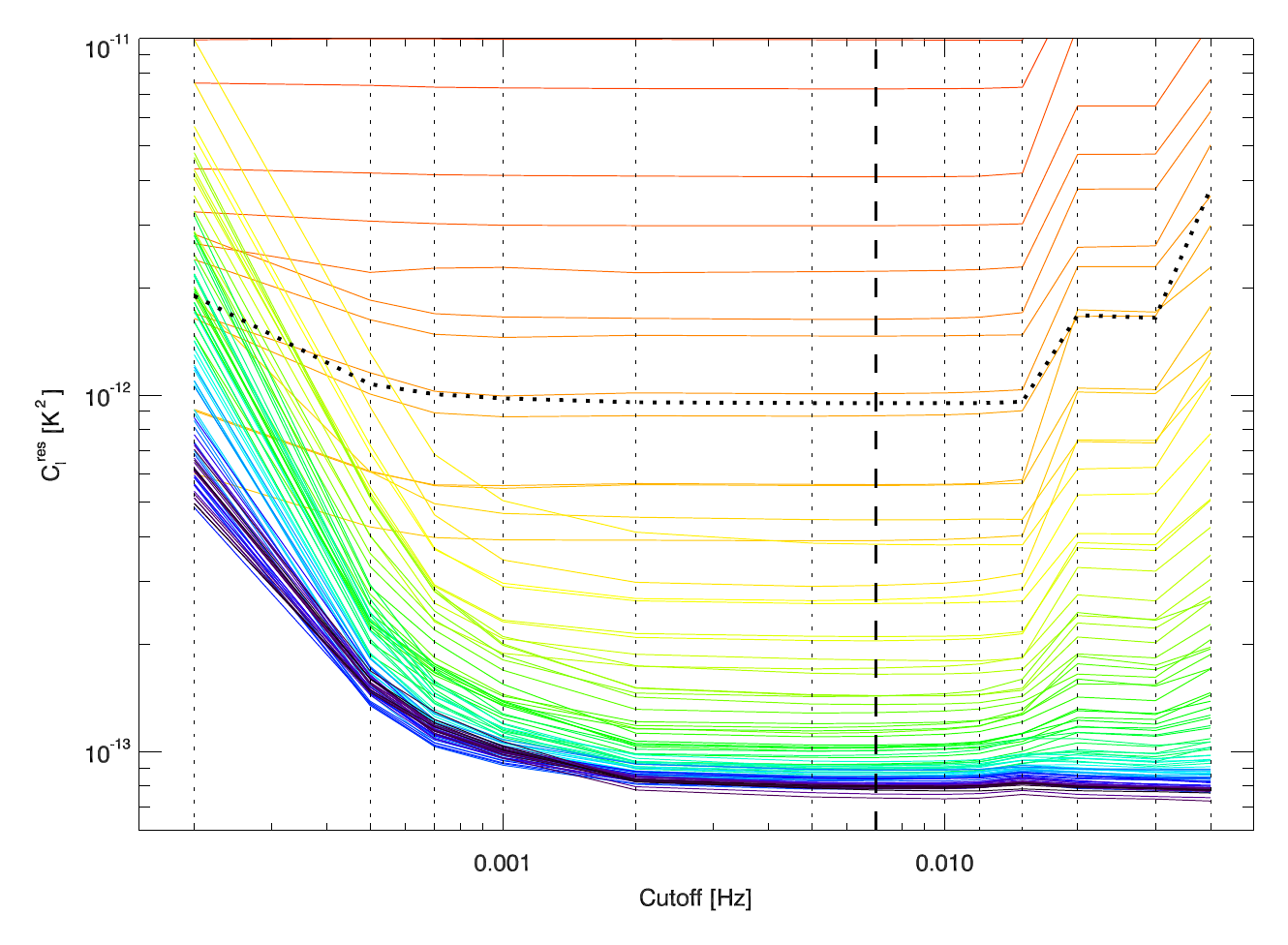}}
            \caption{
                Residual $C_\ell^{\rm res}$ at multipoles ranging from 1 (red) to 500 (blue)
                as a function of the cut-off frequency $f_{\rm cut}$.
                Vertical dotted 
                lines represent the values of $f_{\rm cut}$ that have been tested and the black dashed line is the 
                optimal one. 
                The horizontal dotted line represents the mean of the 500 multipoles. In these runs, we have
                multiplied the amplitude of the periodic systematic effect by a factor 100 to 
                enhance its effect.}
            \label{fig:filt_parameter}
        \end{center}
    \end{figure}

    In Fig.~\ref{fig:filt_parameter}, we plot $C_\ell^{\rm res}$ for various values of the 
    multipole $\ell$ as a function of the cut-off frequency $f_{\rm cut}$. The $n$ and $N$ parameters were
    fixed, in these runs, at the values of $n=24$ and $N=24$~hours, respectively.
    Smaller residuals clearly imply that the filter has a smaller effect on the astrophysical signal and that there is a larger systematic effect removal.

    If the cut-off frequency is too low, then the filter is ineffective in removing the periodic effect
    components at frequencies $\sim~0.17~\unit mHz$ 
    ($\equiv~1/5640\unit~s^{-1}$) and $\sim~1~\unit mHz$ ($\equiv~1/940\unit~s^{-1}$), as reflected
    in the increase of $C_\ell^{\rm res}$ at low frequencies
    
    When $f_{\rm cut}$ approaches the scan frequency, the filter then causes a significant removal
    of the first sky-signal peak, located at $0.016\unit~Hz$ ($\equiv~1/60\unit~s^{-1}$).
    This results in $C_\ell^{\rm res}$ increasing at values of $f_{\rm cut}>0.016$~Hz.
    For even higher values of $f_{\rm cut}$, the 
    filter starts to remove also overtones, causing a second increase in $C_\ell^{\rm res}$.
    An optimal value for the cut frequency has been fixed at $f_{\rm cut} = 0.007\unit Hz$.

    \begin{figure}[t]
            \centering            
            \resizebox{\columnwidth}{!}{\includegraphics[width=9cm]{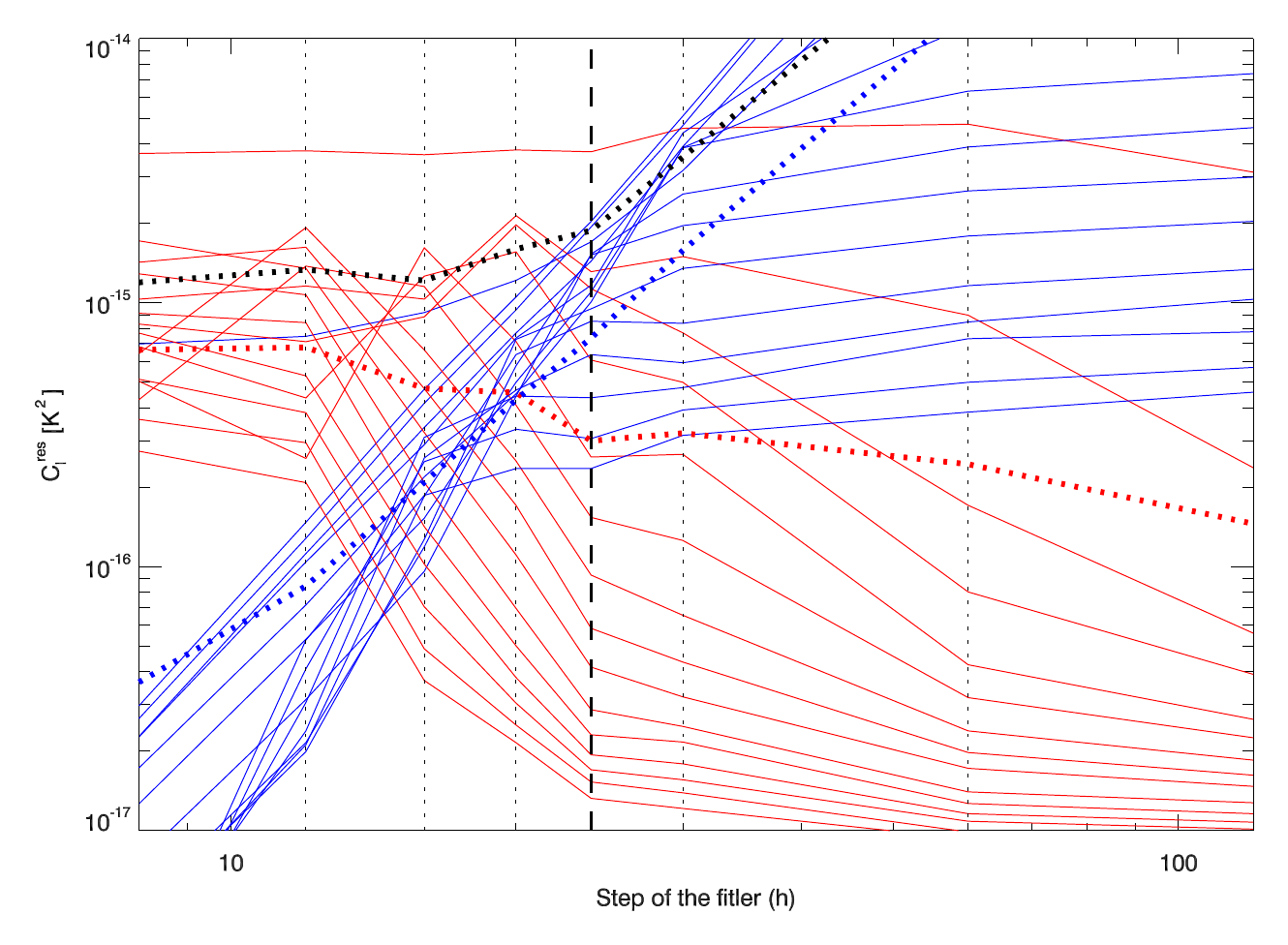}}
            \caption{
                Residual $C_\ell^{\rm res}$ at various multipoles (ranging from 1 to 600 and
                averaged in groups of 40)
                as a function of the filter step, $N$. Red curves represent the residual
                after filtering a data-stream containing only the
                systematic effect, while blue curves show the residual
                obtained after filtering the astrophysical signal. 
                Vertical dotted lines represent the tested values 
                and black dashed line is the selected one. 
                Horizontal coloured dotted lines are the mean of all 600 
                multipoles and the black line represents the sum of the red and blue dotted lines. 
            }
            \label{fig:filt_parameter2}
    \end{figure}

    The effect of the sensitivity analysis on the filter step is shown in Fig.~\ref{fig:filt_parameter2}
    (see caption for details).
    The results show that $N$ must be chosen to be high enough to 
    make the filter capable of detecting the systematic effect but low enough to avoid excessive removal 
    of long-period astrophysical signals. In particular, we see a rather flat minimum in 
    the residual effects up
    to a filter step of $N = 24$ hours, which is the value chosen for our simulations. 

    Another degree of freedom that has been studied is the ``baseline removal''
    i.e. the removal of the zero frequency component
    from the filtered spectrum. In this case, the choice also comes from a trade-off between an effective
    cleaning of the periodic effect (that benefits also from the baseline removal) and the requirement 
    that the filter must not alter the astrophysical signal (which is affected by the baseline removal).
    The overall effect of the two different choices is discussed in Sect.~\ref{sec:results_overview}.
%
%
%
\section{Results and discussion}
\label{sec:results_overview}

  We compare the Fourier filtering (with and without baseline removal) with the destriping process and a combination of the two, in removing the periodic spurious signal shown in the right panel of Fig.~\ref{fig:tods}.

We first compare results obtained with the two different approaches in two simple cases: (i)~a data stream containing only the periodic effect and (ii)~a data stream containing only the astrophysical signal. We then perform the same comparison by also taking into account the $1/f$ noise.

\subsection{Filter applied to periodic signal alone}
\label{sec_periodic_signal_alone}

    \begin{figure*}[!ht]
        \centering
        \footnotesize
        \def\arraystretch{1.5}
        \begin{tabular}{x{2cm}x{2.5cm}x{2.5cm}x{2.5cm}x{2.5cm}x{2.5cm}}
            \hline\hline
            Original noise & 
            After Destrip & 
            After HPF & 
            After HPF and Destriping & 
            After HPF (baseline removed) & 
            After HPF (baseline removed) and Destriping\tabularnewline
            \hline
            $\approx 6.30\unit \mu K$ &
            $\approx 0.95\unit \mu K$ &
            $\approx 1.87\unit \mu K$ &
            $\approx 0.86\unit \mu K$ &
            $\approx 0.84\unit \mu K$ &
            $\approx 0.86\unit \mu K$ \tabularnewline
            &(damping $\sim 6.6$) &
            (damping $\sim 3.4$) &
            (damping $\sim 7.3$) &
            (damping $\sim 7.5$) &
            (damping $\sim 7.3$)
            \tabularnewline
            \hline\hline
        \end{tabular}
        \subfloat[Periodic systematic effect]{
            \includegraphics[width=0.4\textwidth, angle=180]{figures/map256-staticR_cooler_LFI-27a_1survey.pdf}
            \label{fig:map256-staticR_cooler_LFI-27a_1survey-2}}
        \subfloat[After destriping]{
            \includegraphics[width=0.4\textwidth, angle=180]{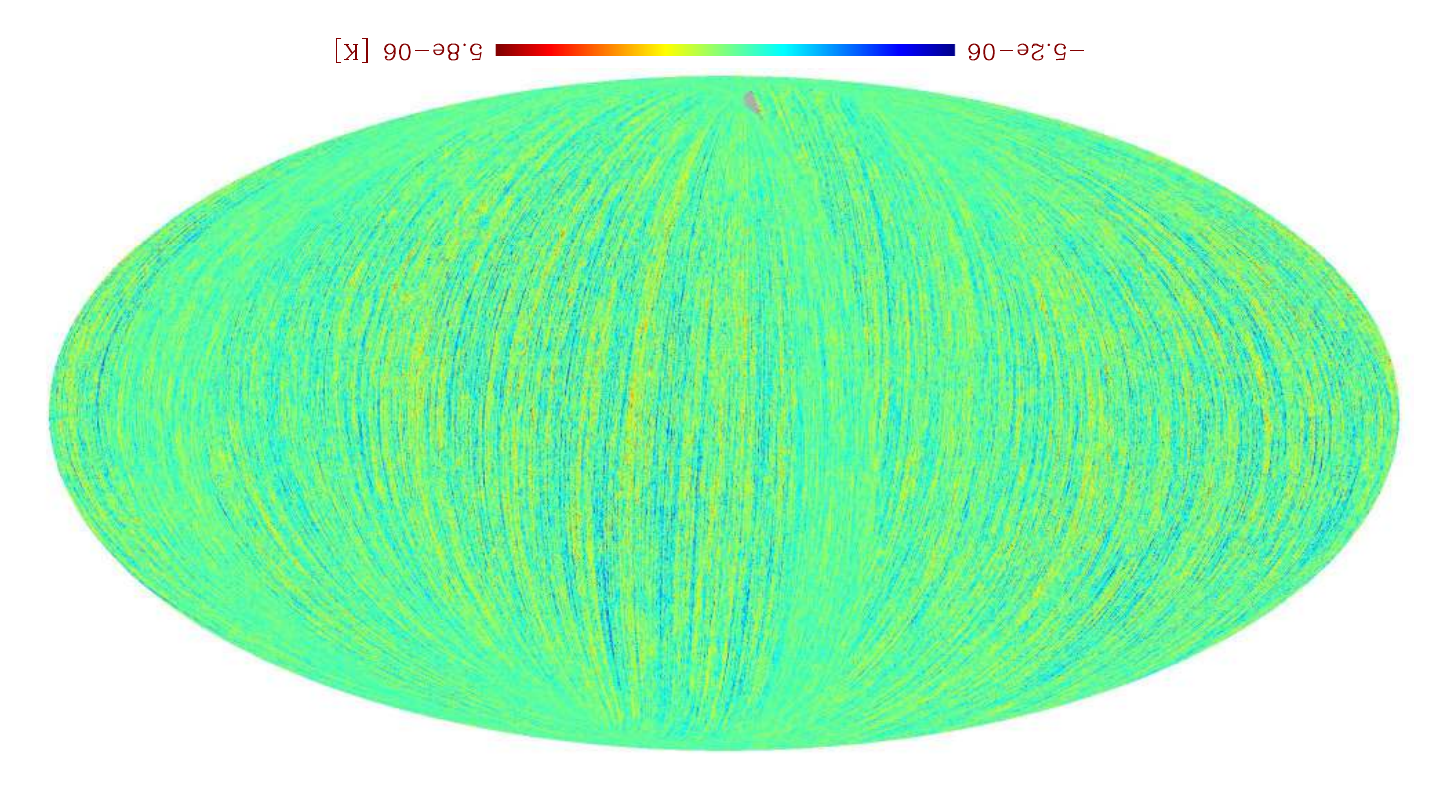}
            \label{fig:map256-destrip-staticR_cooler_LFI-27a_1survey}}\\
        \subfloat[After filtering]{
            \includegraphics[width=0.4\textwidth, angle=180]{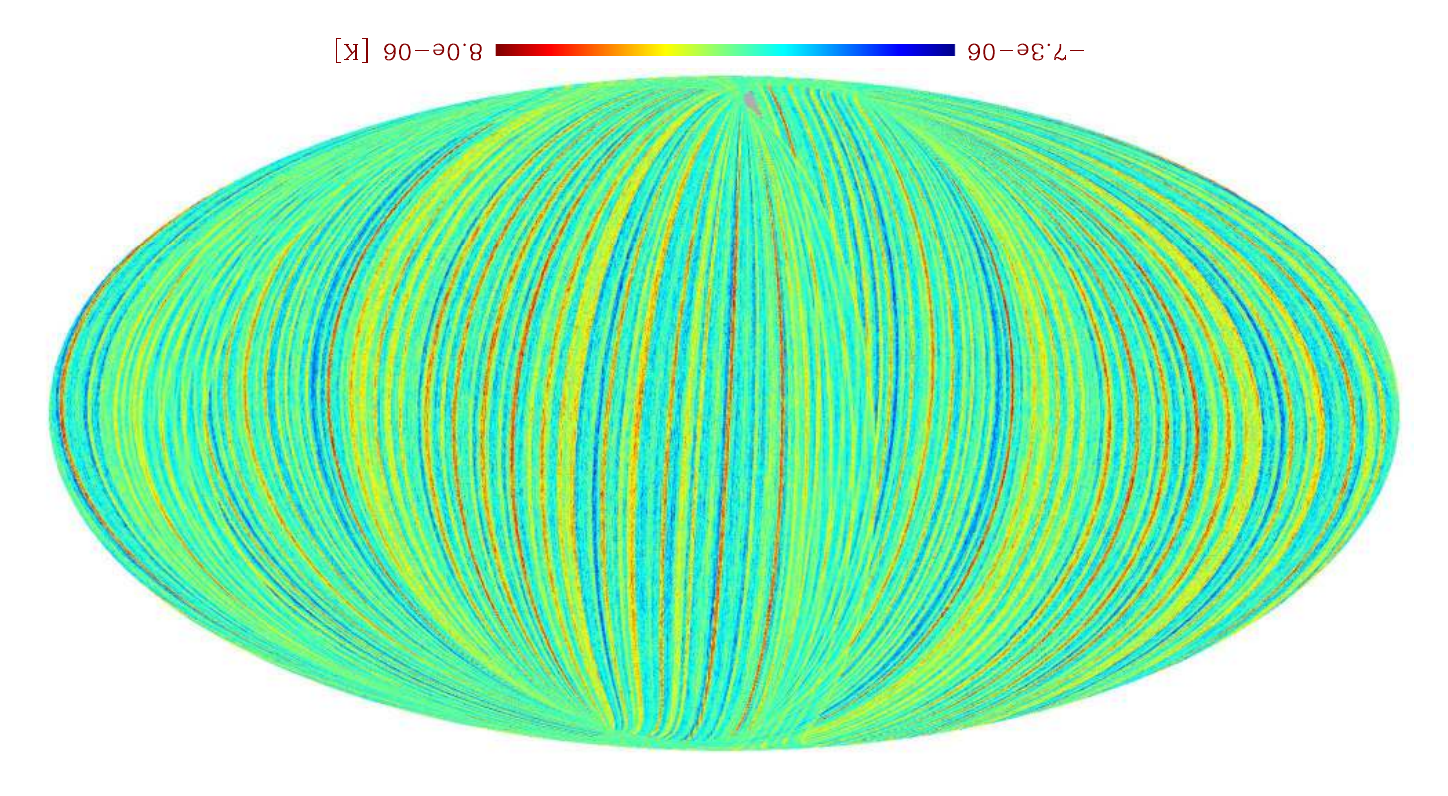}
            \label{fig:map256-filt-staticR_cooler_LFI-27a_1survey}}
        \subfloat[After destriping and filtering]{
            \includegraphics[width=0.4\textwidth, angle=180]{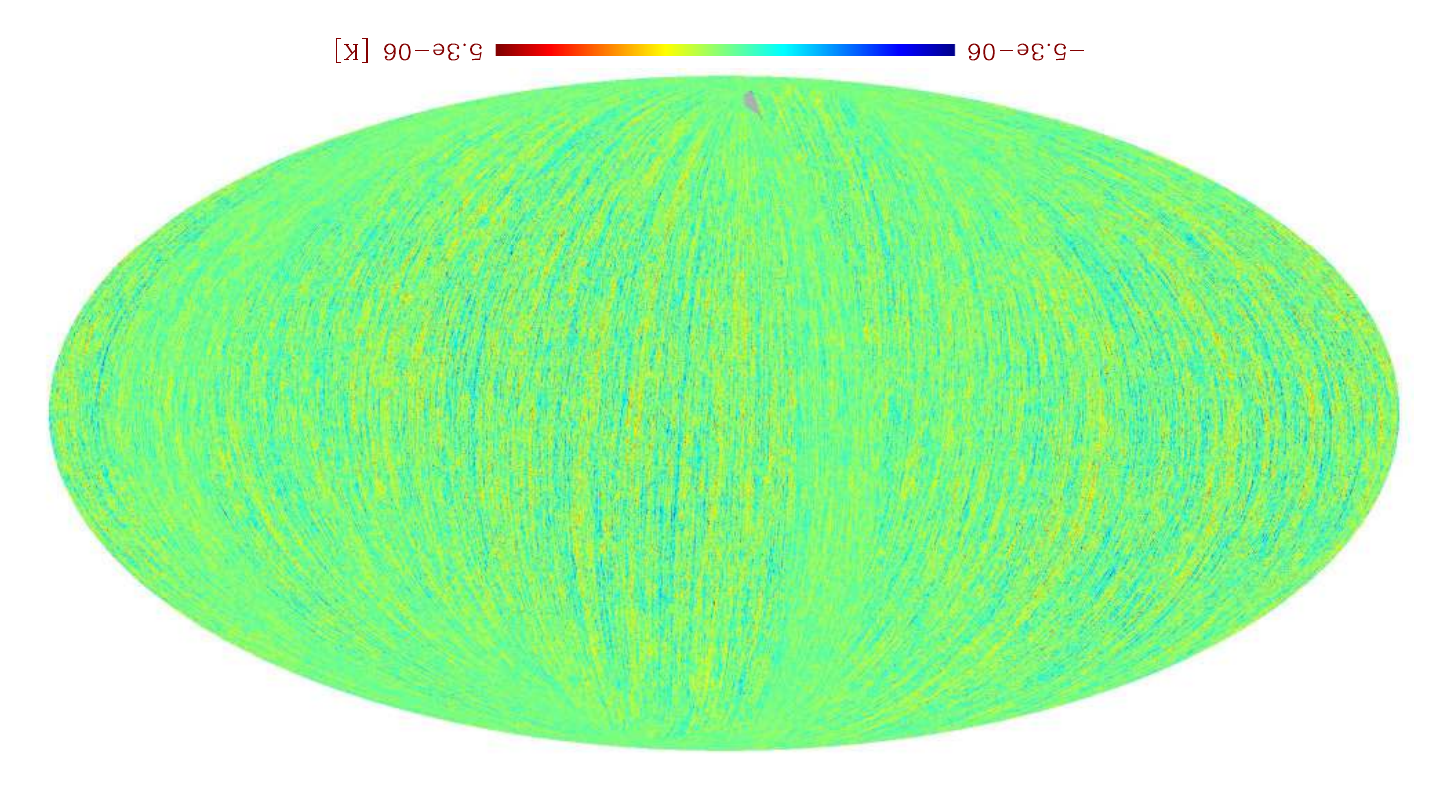}
            \label{fig:map256-destrip-filt-staticR_cooler_LFI-27a_1survey}}\\
        \subfloat[After filtering (baselines removed)]{
            \includegraphics[width=0.4\textwidth, angle=180]{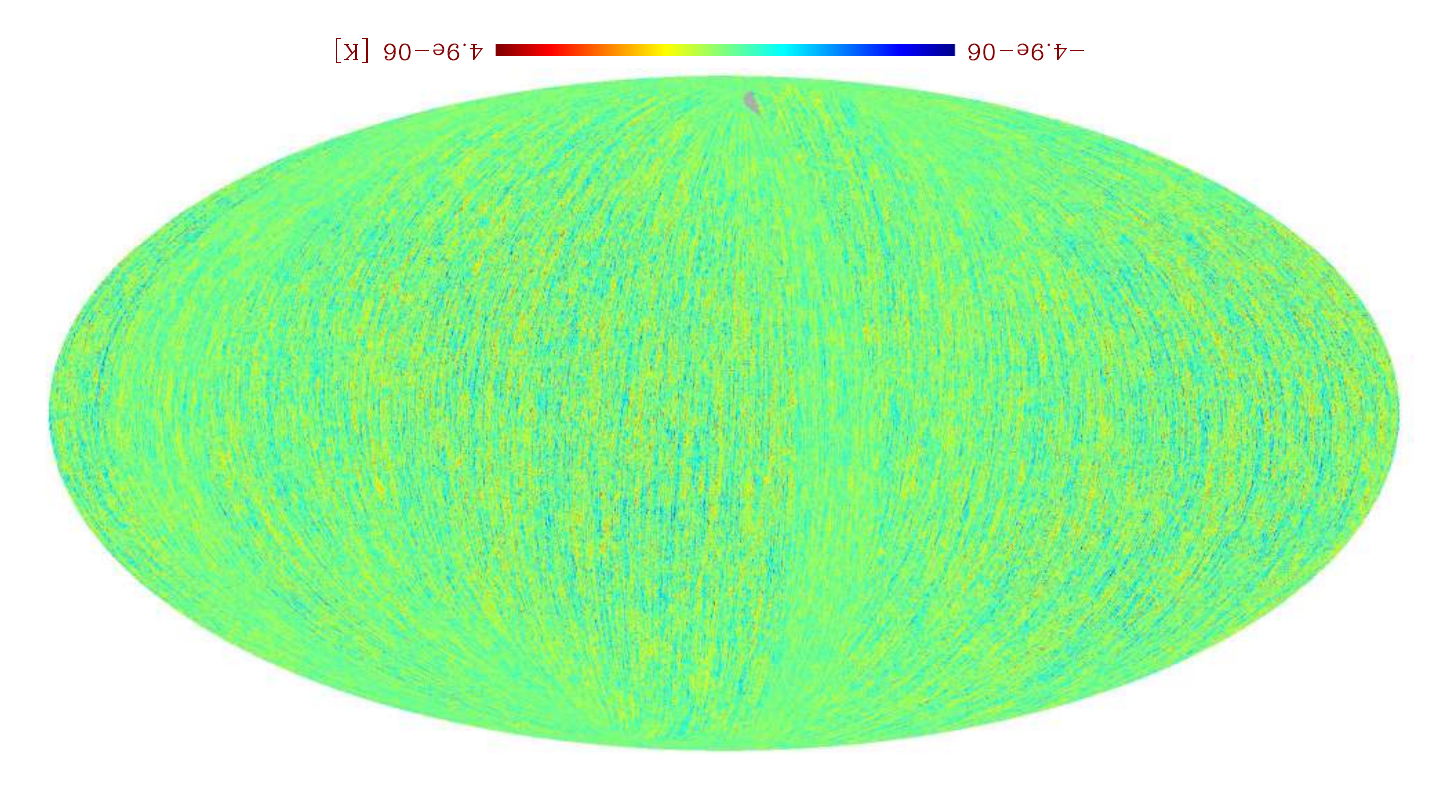}
            \label{fig:map256-filtB-staticR_cooler_LFI-27a_1survey}}
        \subfloat[After destriping and filtering (baselines removed)]{
            \includegraphics[width=0.4\textwidth, angle=180]{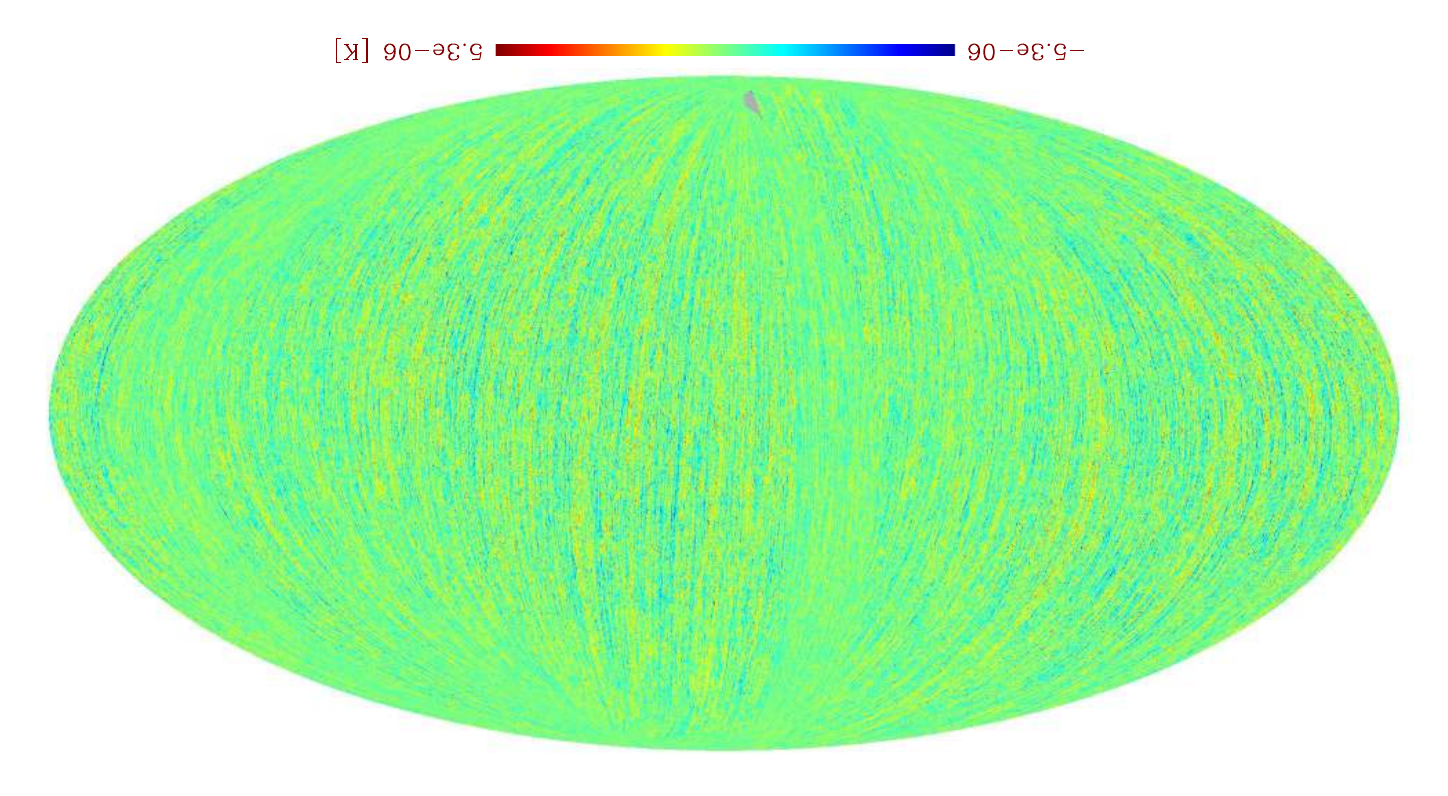}
            \label{fig:map256-destrip-filtB-staticR_cooler_LFI-27a_1survey}}
        \caption{
            Maps of the periodic systematic effect before and
            after filtering-alone (with and without baseline removal), 
            destriping-alone, and filtering + destriping. Colours are rescaled in every map to highlight footprints of any systematic effects. In the above table, we summarise the map r.m.s. values and the corresponding
	    dumping factors.
        }
        \label{fig:filter_destrip_cooler}
    \end{figure*}

    In Fig.~\ref{fig:filter_destrip_cooler}, the periodic systematic effect is projected onto a HEALPix with $N_{\rm side} = 256$ (corresponding to a pixel size of $\sim 13.7$~arcmin) before and after the application of various removal algorithms. The peak-to-peak effect on the map is $\approx 66\unit \mu K$, that is reduced to $\approx 11\unit \mu K$ after filtering and destriping.
    
    If we now consider the results obtained by filtering without destriping (maps (c) and (e)) we can see that:
    \begin{itemize}
        \item when the baselines are not removed 
        (Fig.~\ref{fig:map256-filt-staticR_cooler_LFI-27a_1survey}), the residual map displays 
        stripes caused by long-period oscillations that are not removed by the filter. These stripes
        can be removed by a subsequent application of the destriping code;
        \item when baselines are removed, the filtered-only map 
        (Fig.~\ref{fig:map256-filtB-staticR_cooler_LFI-27a_1survey}) is similar to the destriped-only
        map (Fig.~\ref{fig:map256-destrip-staticR_cooler_LFI-27a_1survey}). Additional 
        destriping after filtering (Fig.~\ref{fig:map256-destrip-filtB-staticR_cooler_LFI-27a_1survey})
        does not yield any measurable improvement.
    \end{itemize}

    \begin{figure}[!tb]
        \centering
        \resizebox{\columnwidth}{!}{\includegraphics{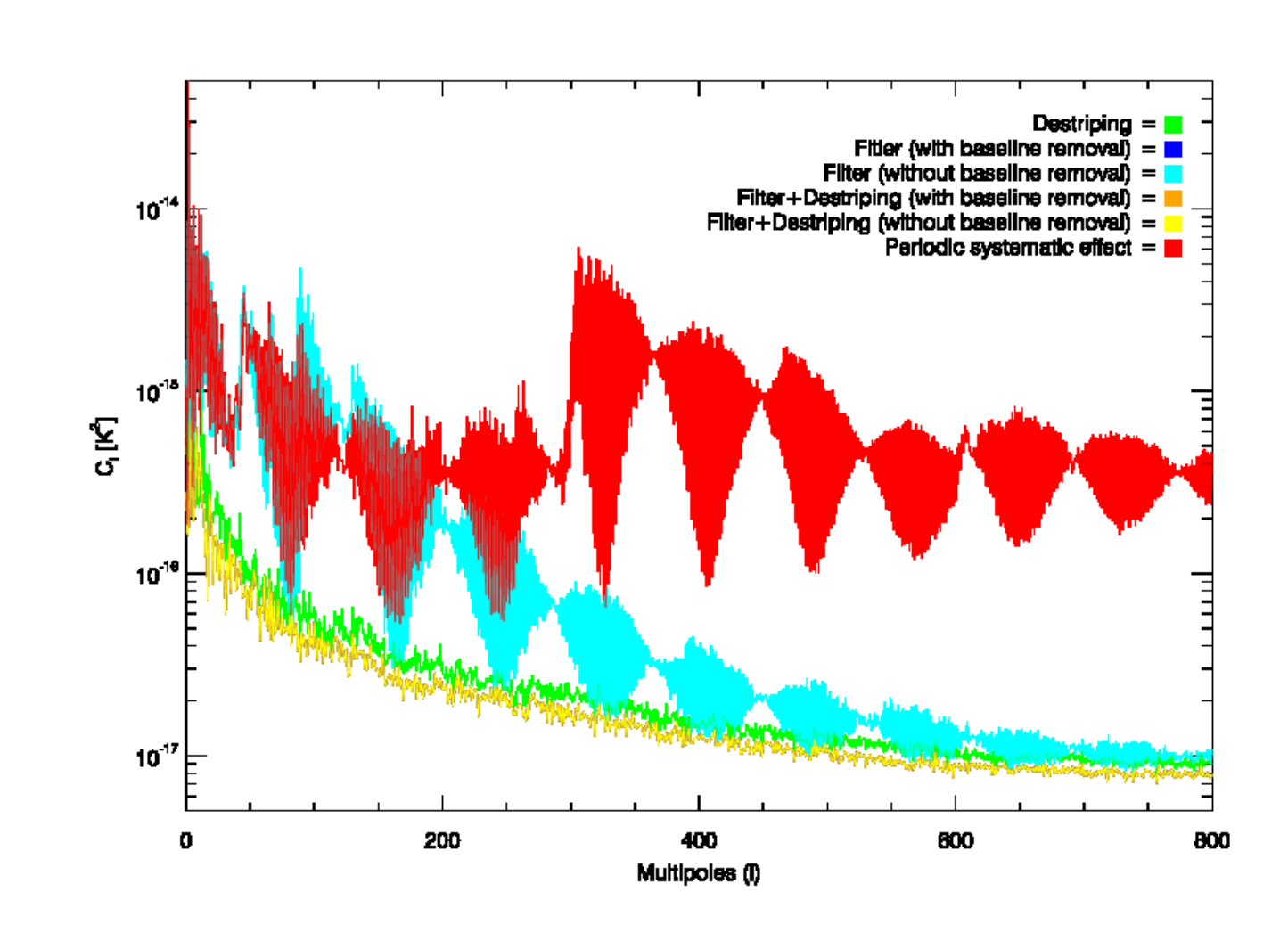}}
        \resizebox{0.95\columnwidth}{!}{\includegraphics[width=9cm]{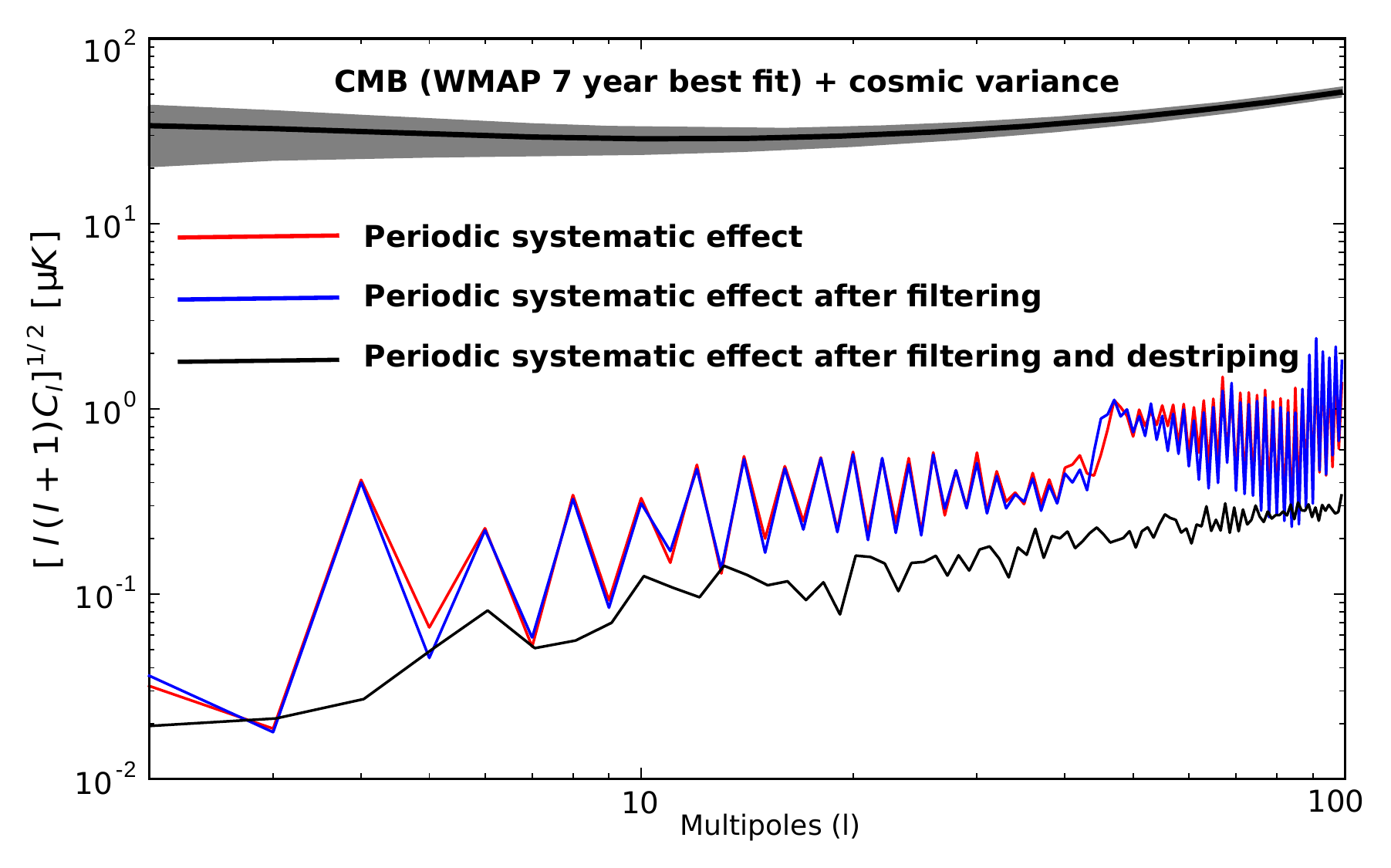}}
        \caption{
            Top: Power spectra the periodic systematic effect before (red curve) and after removal.
            Green: destriping only. Filtering without baseline removal -- 
            Cyan: filtering only, yellow: filtering + destriping. 
            Filtering with baseline removal -- 
            Blue: filtering only, orange: filtering + destriping.
	    Note that
	    the yellow, orange, and blue curves are superimposed, meaning that the three cases
            produced the same results.
	    Bottom: Impact of periodic fluctuations on large scales compared to the CMB and cosmic variance obtained from a best-fit solution of the seven year WMAP data.
        }
        \label{fig:spettri_cooler}
    \end{figure}

    To study the direct impact of the periodic signal on the CMB, we compare power spectra as done in the top panel of Fig.~\ref{fig:spettri_cooler}. From the figure we see that best performance is obtained with the combination of filter and destriping. With the filter-only, in particular, baseline removal plays a key role, as it leaves stripes that otherwise need to be subsequently removed by destriping. If we use the filter + destriping combination, however, baseline removal becomes irrelevant and the residual after the combined cleaning procedure is of the order of $\sim 70\%$ of the residual obtained by applying destriping only.
 
     In the bottom panel of Fig.~\ref{fig:spettri_cooler}, we compare the CMB power spectrum obtained from the seven year WMAP best fit \citep{jarosik2010} including cosmic variance with the residual caused by the periodic effect. The figure shows that in all cases the effect is at least two decades below the CMB spectrum and that the combination of filtering and destriping is able to reduce the residual effect by another order of magnitude. Similar comparisons at small angular scales are provided in Fig.~\ref{fig:final}.

\subsection{Filter applied to astrophysical signal alone}

    In the next step we applied the same algorithms to a data stream containing only a simulated astrophysical signal composed by CMB + galactic diffuse emission.
    
    \begin{figure}[!tb]
        \resizebox{\columnwidth}{!}{\includegraphics[angle=180]{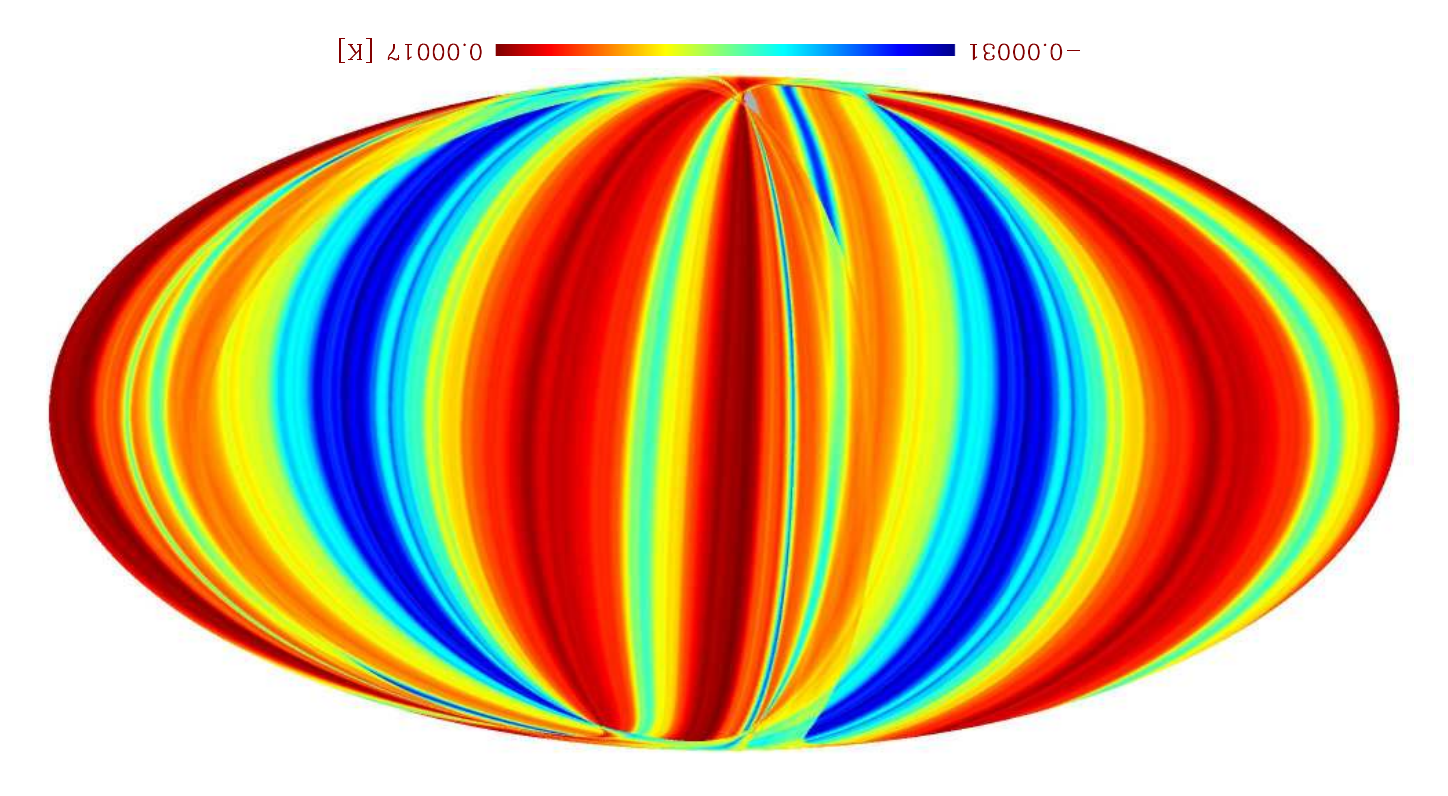}
        }
        \resizebox{\columnwidth}{!}{\includegraphics[angle=180]{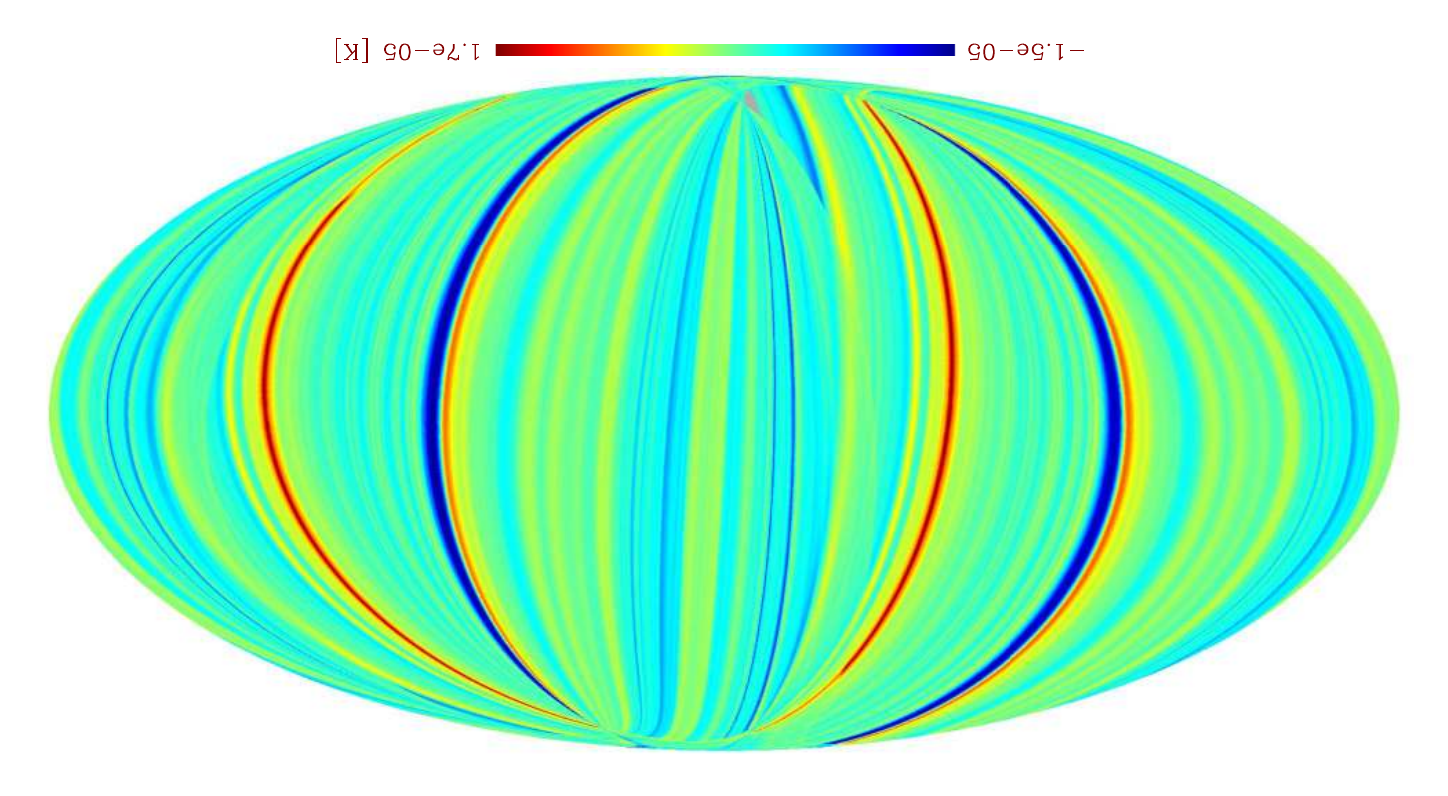}
        }
        \resizebox{\columnwidth}{!}{\includegraphics{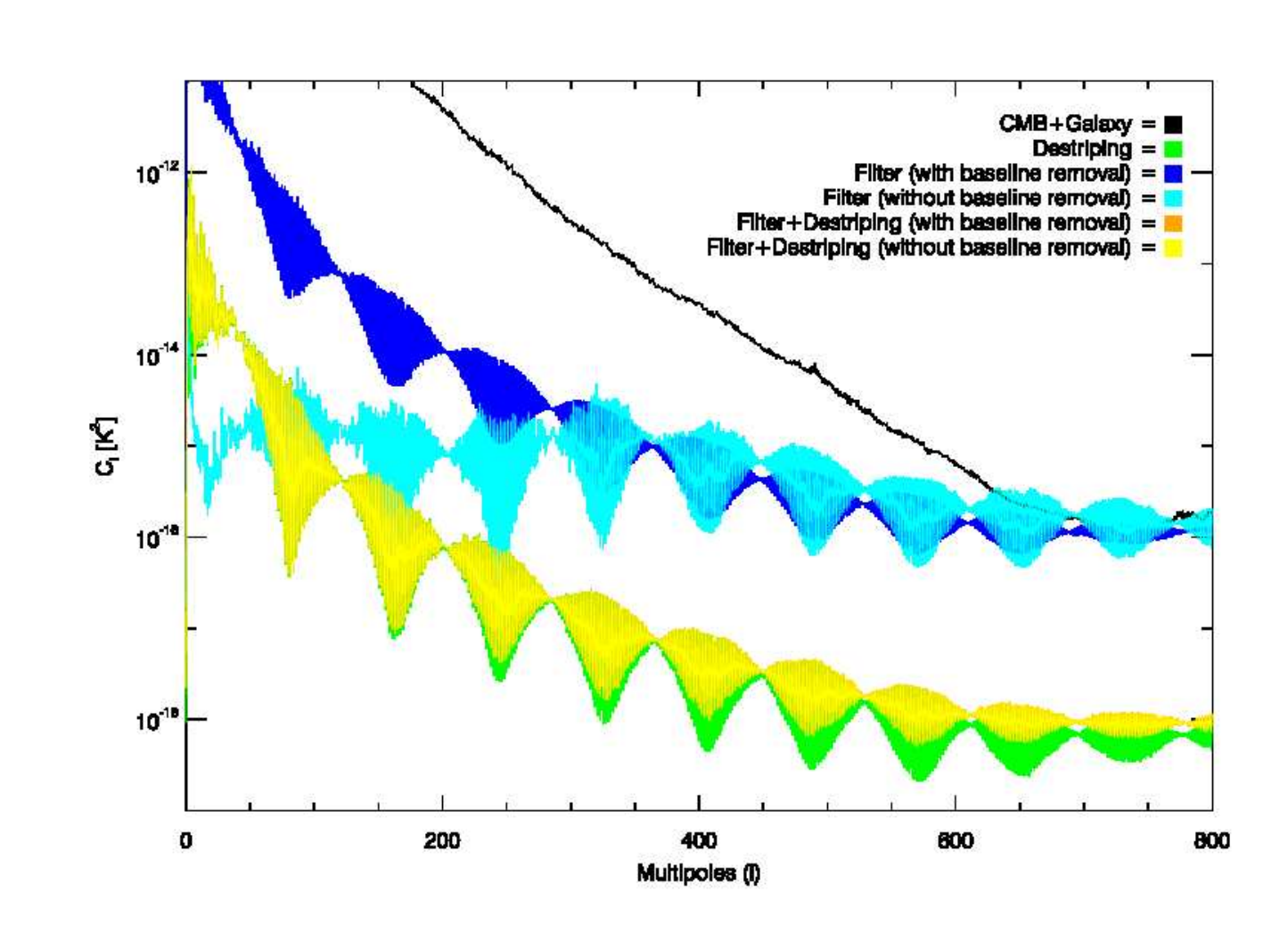}}
        \caption{Top: residual CMB+Galaxy map after filtering with baseline removal. 
            Middle: residual CMB+Galaxy map after destriping. 
            Bottom: power spectra of residual CMB+galaxy maps after destriping (green), filter-only without
            baseline removal (cyan), filter-only with baseline removal (blue), filter without baseline
            removal + destriping (yellow), filter with baseline
            removal + destriping (orange, below yellow curve).
         }
    \label{fig:stripes}
    \end{figure}

    In Fig.~\ref{fig:stripes}, we show the power spectra of residual maps obtained after applying of the various filtering approaches. Residual maps are defined as
    
    \begin{equation}\label{eq:residual-sky}
        M_{\rm(residual)} = \left[ M_{\rm(sky)}\right]_{\rm HPF/Dest} - M_{\rm(sky)},
    \end{equation}    
    where $M_{\rm(sky)}$ is the reference sky map and $\left[ M_{\rm(sky)}\right]_{\rm HPF/Dest}$ is the same map after the application of the cleaning code (with any of the tested combinations).

    The curves in Fig.~\ref{fig:stripes} tell us that any kind of algorithm applied to clean systematic effects is going to also alter the astrophysical signal. The high-pass filter, in particular, leaves a residual especially when baselines are removed. This is probably due to the strong difference in the mean value of the various chunks of data because the signal is mostly concentrated along some rings (blue patches in the top panel of Fig.~\ref{fig:stripes}) and, therefore, in just some data-chunks. When the mean values of these data-chunks are removed, strong marked stripes are generated in the residual map. From these results, it is clear that any application of Fourier filters to real data requires additional baseline normalisation such as that performed by destriping codes.
    
\subsection{Filter applied to the combined signal}

    \begin{figure*}[!tb]
    \centering    
    \subfloat[Signals]{
        \includegraphics[width=0.4\textwidth]{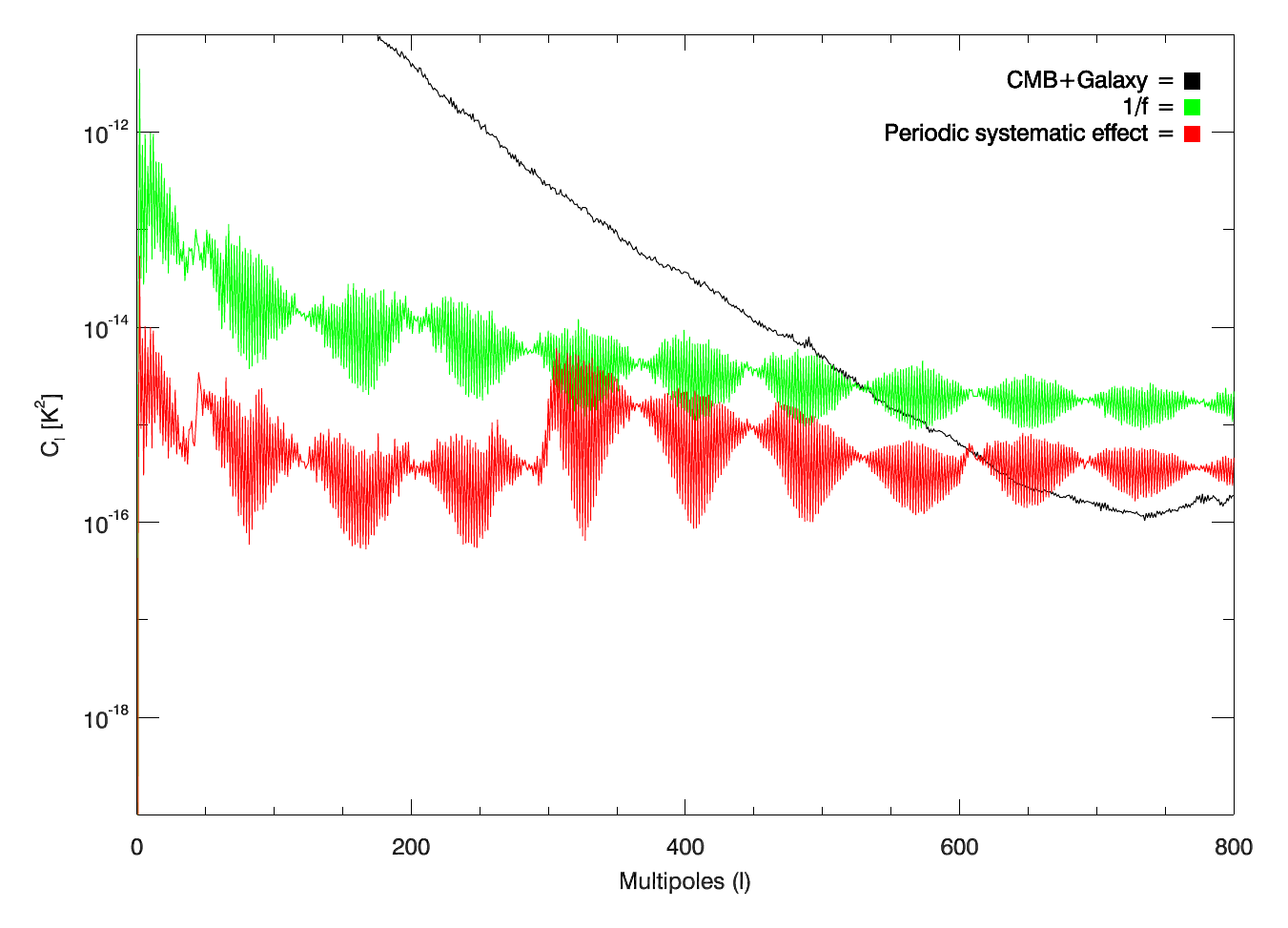}
        \label{fig:spettro_segnali}}
    \subfloat[Destriper effect]{
        \includegraphics[width=0.4\textwidth]{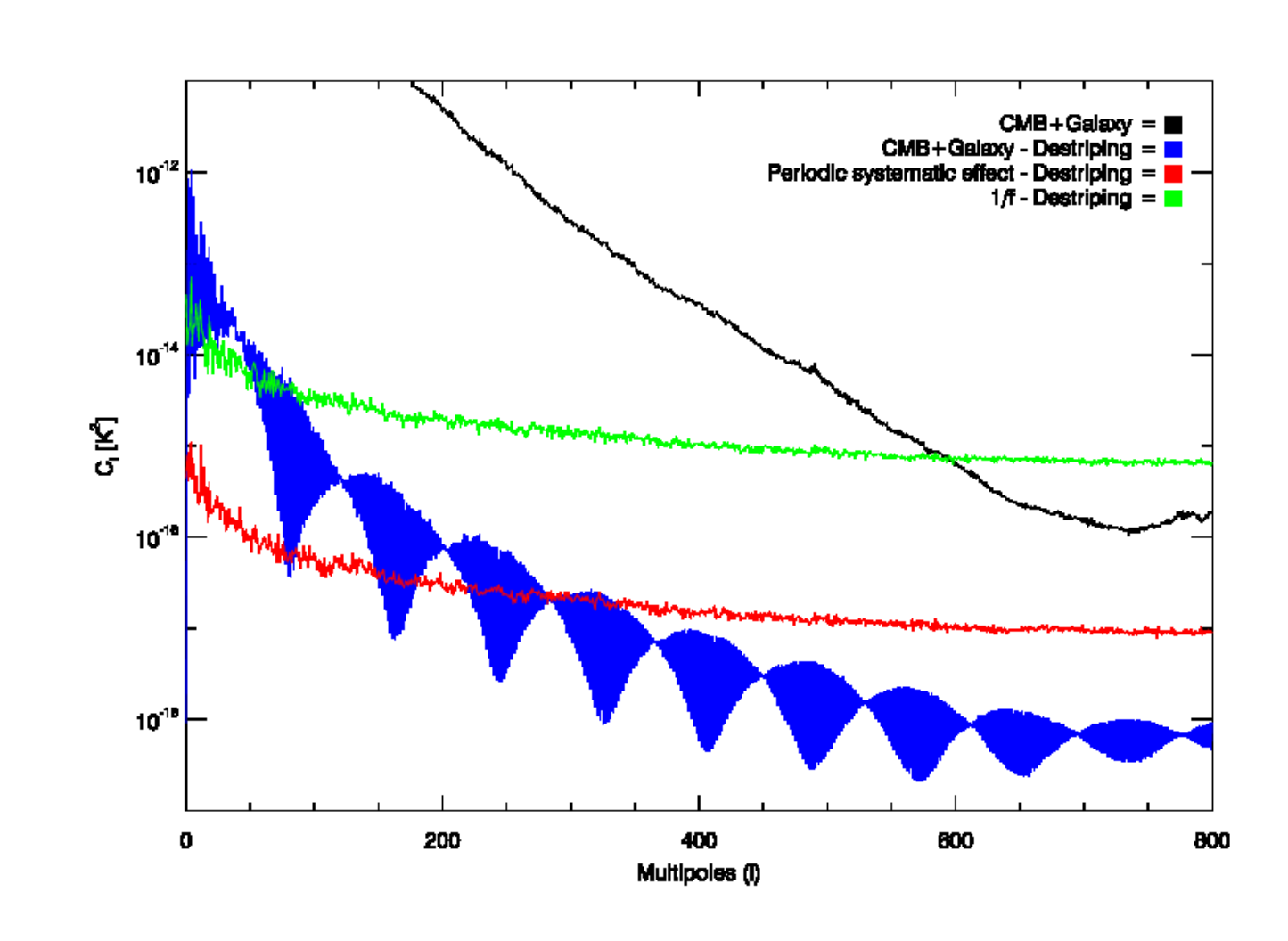}
        \label{fig:spettro_final-destrip}}\\
    \subfloat[Filter effect]{
        \includegraphics[width=0.4\textwidth]{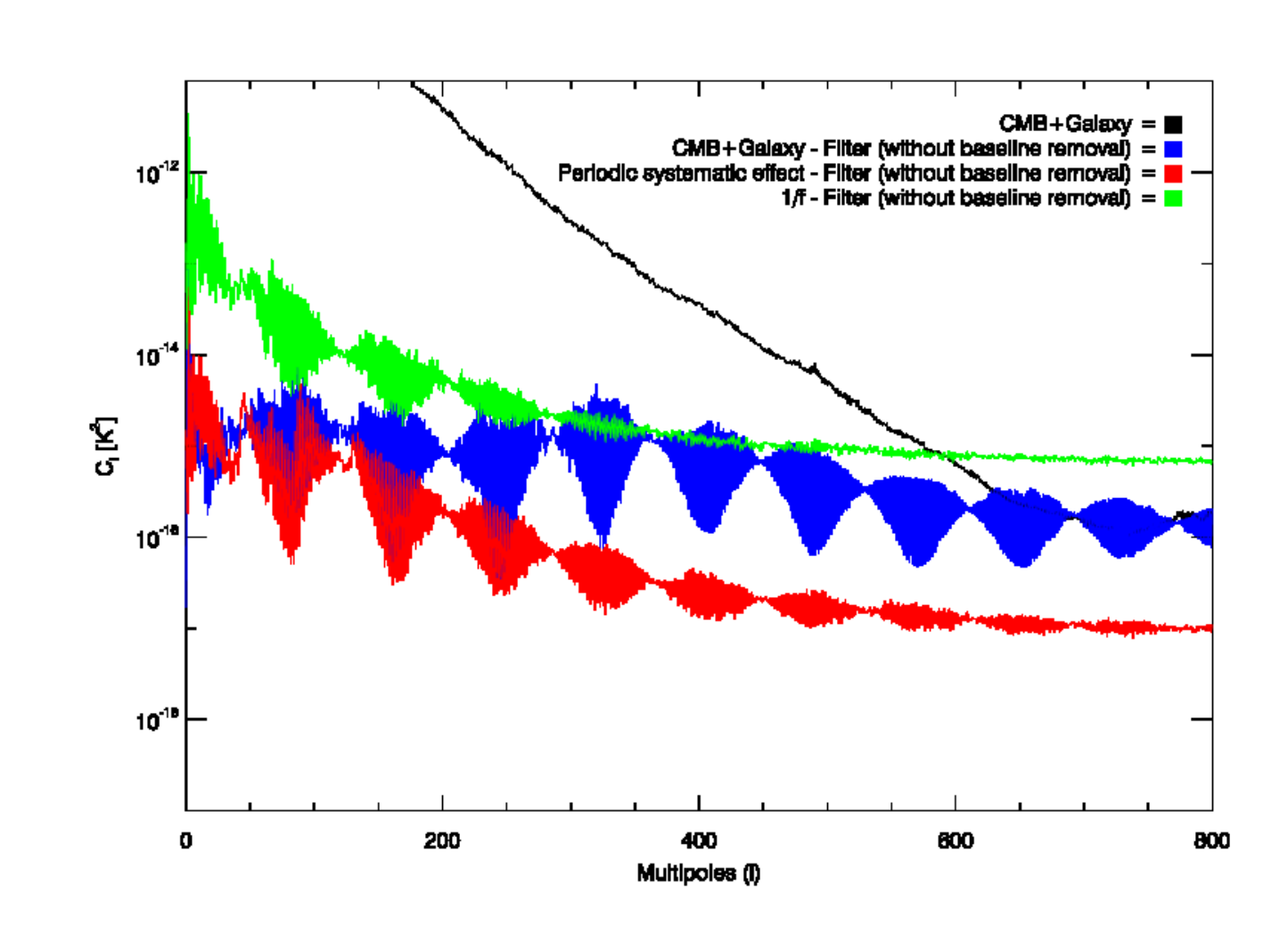}
        \label{fig:spettro_final-filt}}
    \subfloat[Filter + Destriper effect]{
        \includegraphics[width=0.4\textwidth]{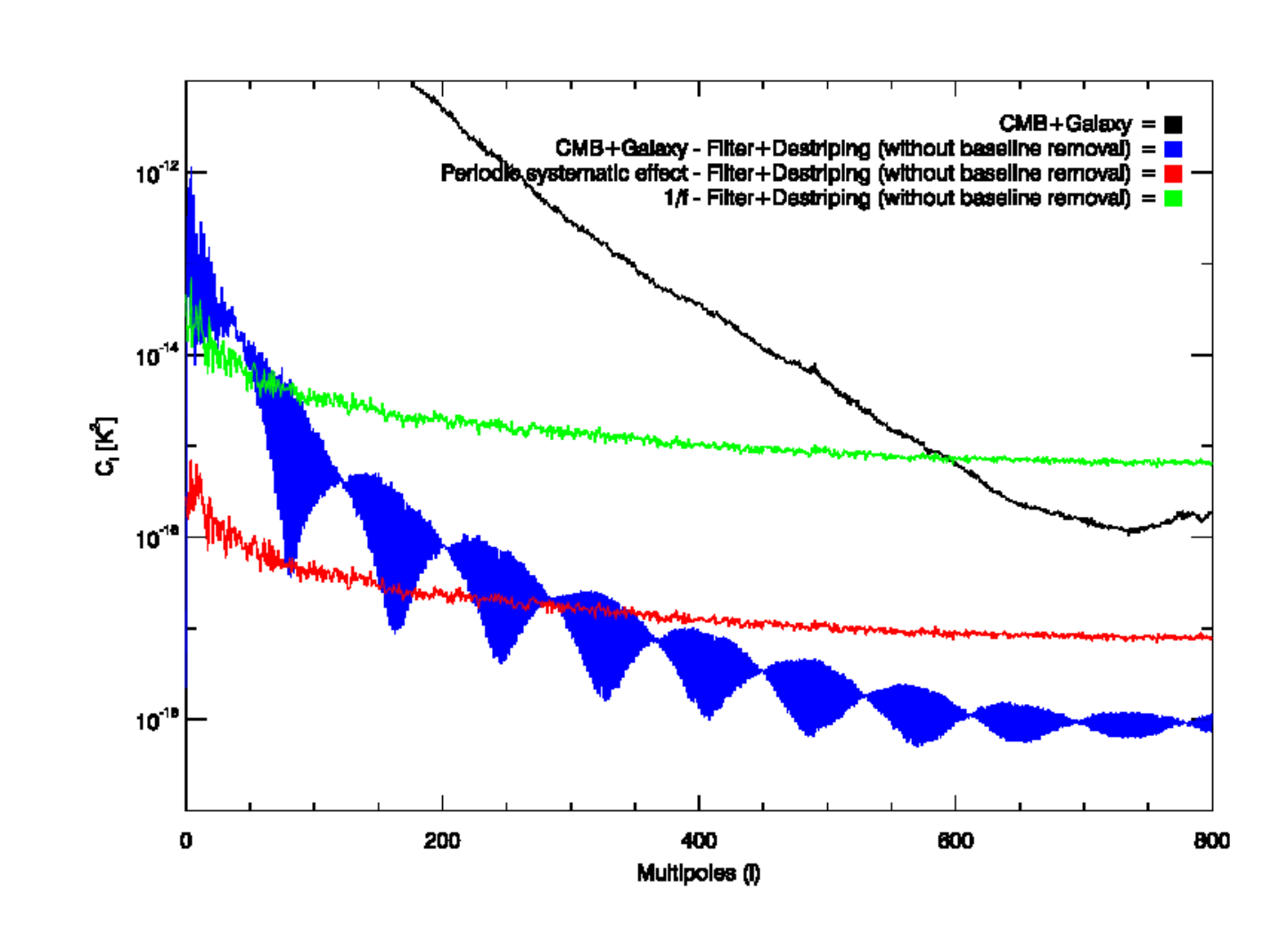}
        \label{fig:spettro_final-filt+destrip}}\\
    \subfloat[Filter (no baselines) effect]{
        \includegraphics[width=0.4\textwidth]{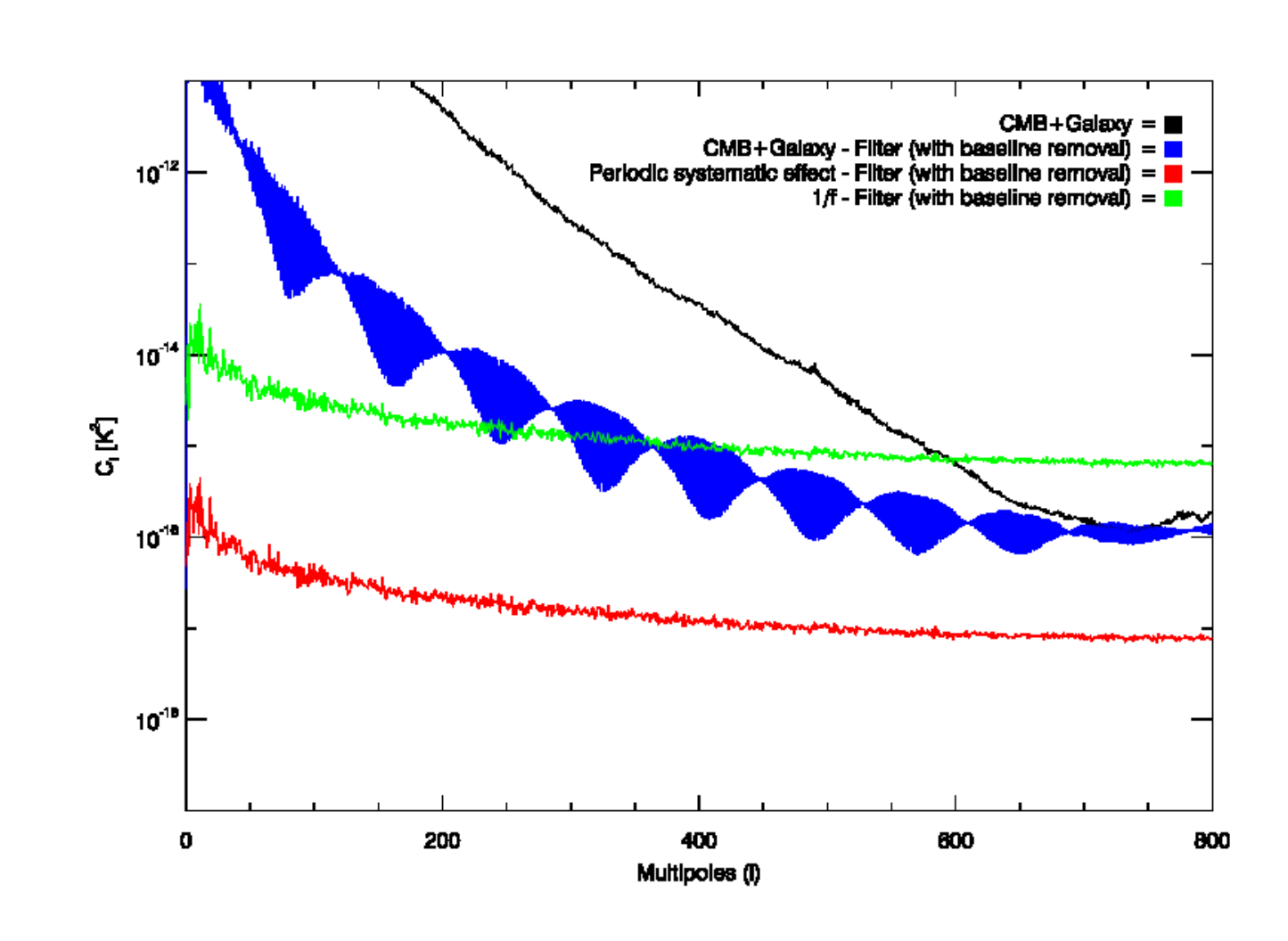}
        \label{fig:spettro_final-filtb}}
    \subfloat[Filter (no baselines) + Destriper effect]{
        \includegraphics[width=0.4\textwidth]{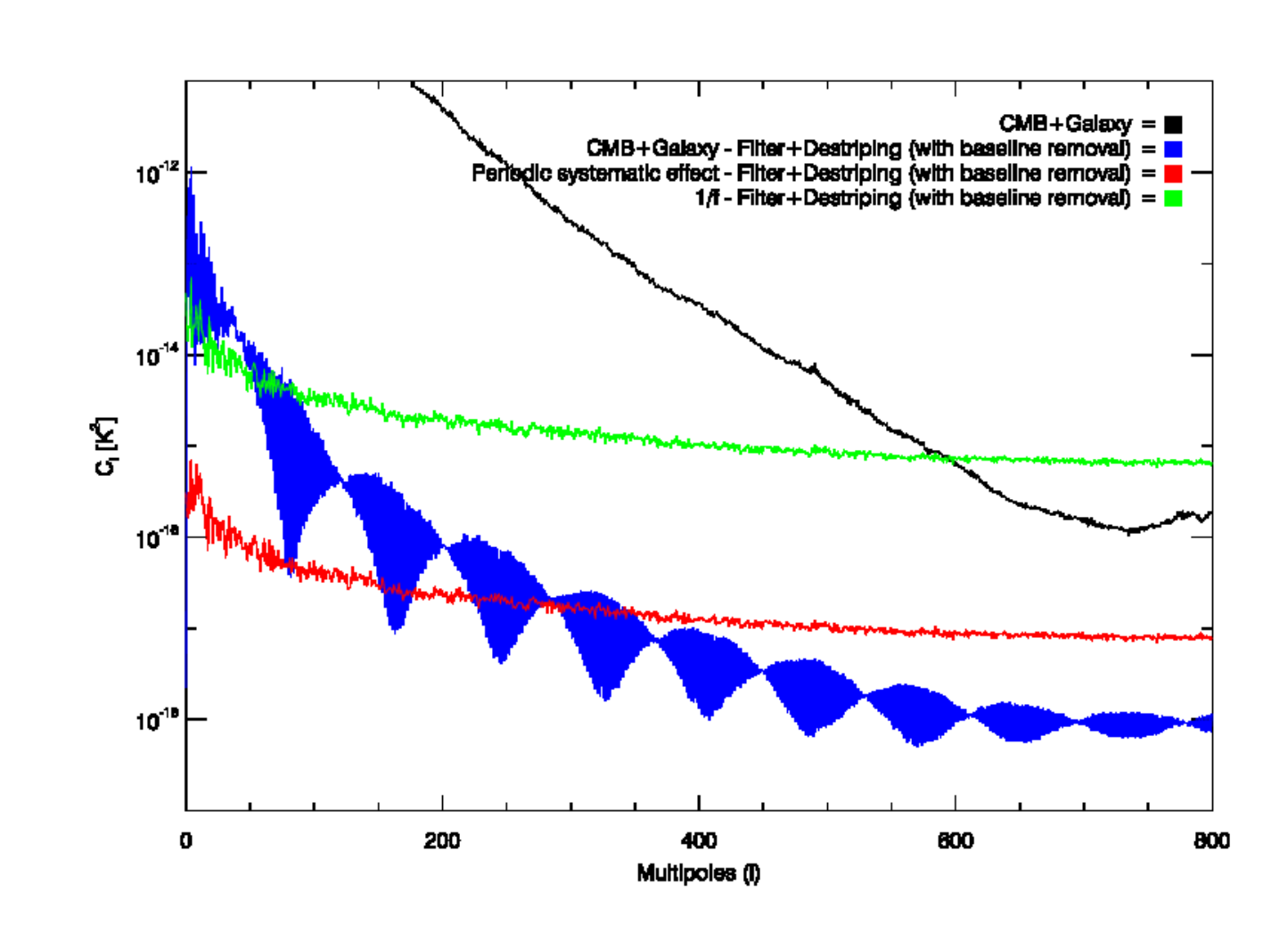}
        \label{fig:spettro_final-filtb+destrip}}
    \caption{Power spectra main components before (a) and after 
    cleaning with destriping (b), high-pass filter (c) and code-combination (d). 
    The last line shows the effect when the baselines are removed by high-pass filter (e) and 
    the code-combination (f).}
    \label{fig:final}
    \end{figure*}

    The third step was to apply various cleaning procedures to data-streams containing the sky signal, the 1/$f$ component, and the periodic systematic effect and then compare the results. Since all the steps (filtering, destriping/mapmaking, and spectrum extraction) are linear, we can work separately on the different
    components without any loss of information.

    In Fig.~\ref{fig:final}, we plot the residual power spectra for every signal separated by cleaning method. Fig.~\ref{fig:spettro_segnali} shows the power spectra of the various signals when no cleaning is applied. In the other panels, we report the same spectra after applying various cleaning techniques together with the residual coming from the modification of the sky signal caused by the software removal. The green curve in Fig.~\ref{fig:spettro_final-destrip}, in particular, shows the residual level from $1/f$ noise after destriping. Our results indicate that both destriping alone and the combination of filter and destriping can suppress the effect of the periodic signal at a level that is at least two order of magnitudes below this residual.

    Fig.~\ref{fig:spettro_final-filt} (high-pass filter) again highlights the need to use a destriper in order to renormalise offsets generated by the filter. A comparison of Figs.~\ref{fig:spettro_final-destrip} (destriping) and~\ref{fig:spettro_final-filt+destrip} (code-combination) shows that the damping of the systematic effect obtained by applying destriping only is essentially the same as that obtained by combining the two codes.
    
    The two bottom plots show the effects of the high-pass filter and the code-combination with baselines removal. The filter-only baseline removal again produces a larger residual, while no significant difference is seen when the filter is combined with destriping.
%
%
%
\section{Conclusions}
\label{sec:conclusion_further_work}

  We have analysed the use of a high-pass Fourier filter to clean full-sky CMB datasets from periodic systematic effects. As a test case, we have considered the baseline \planck\ satellite scanning strategy and instrument parameters typical of the \planck-LFI 30~GHz receivers. A template of the spectral content of periodic spurious signal was derived from ground tests of the 20~K \planck-LFI focal plane unit. The effectiveness of Fourier filtering was then compared with 
destriping and with a third approach that combines filtering and destriping in sequence.

After a sensitivity study to define the optimal filter parameters, we first processed a data stream containing only the systematic effect. The high-pass filter with baseline removal is effective in removing the periodic signal, with residual effects lower than those obtained by destriping.

We then processed a data stream containing only the astrophysical signal; in this case the filter creates a new artefact, visible in the power spectrum at low multipoles and in the residual maps as sharp stripes. This residual effect can be effectively removed by a subsequent application of destriping.

The final power spectrum (with all the components described in Sec.~\ref{sec:time_streams}) obtained after filter + destriping differs from the one obtained after destriping by $< 10^{-16} {\rm K}^2$ for $\ell < 500$ and $< 10^{-18} {\rm K}^2$ for $\ell > 500$. This means that destriping is able to reproduce almost the same effect obtained by the filter + destriping combination. These results indicate that high-pass Fourier filters are not suitable for cleaning large CMB datasets from periodic systematic effects, especially when destriping can be effectively applied.


\begin{acknowledgements}
The work in this paper has been supported
by in the framework of the ASI-E2 phase of the Planck contract and
has been carried out at the Planck-LFI Data Processing Centre located
in Trieste at the Astronomical Observatory. 
Some of the results in this paper have been derived using the 
HEALPix package \citep{gorski_2005}.
\end{acknowledgements}


\bibliographystyle{aa}
\bibliography{references}

\end{document}